\documentclass[11pt]{article}
\usepackage{emulateapj}

\usepackage{flushrt}
\usepackage{natbib}
\usepackage{journals}      
\usepackage{epsfig}

\def\Msun{\ifmmode {\rm M}_{\mathord\odot}\else $M_{\mathord\odot}$\fi}
\newcommand{\msun}{M_\odot}
\newcommand{\pc}{{\rm pc}}
\newcommand{\tff}{t_{\rm ff}}
\newcommand{\Mej}{M_{\rm ej}}
\newcommand{\mstar}{m_\star}
\newcommand{\vesc}{v_{\rm esc}}

\newcommand{\Mstar}{{M_\star}}
\newcommand{\Mcl}{M_{\rm cl}}

\newcommand{\SFE}{{\rm SFE}}

\newcommand{\SFEtot}{{\rm SFE_{\rm tot}}}
\newcommand{\pw}{p_{w}}
\newcommand{\kms}{{\rm km\;s^{-1}}}
\newcommand{\vw}{v_{w}}
\newcommand{\mw}{m_{w}}
\newcommand{\fw}{f_{w}}
\newcommand{\tw}{t_{w}}
\newcommand	{\simlt}{\lower.5ex\hbox{$\; \buildrel < \over \sim \;$}}
\newcommand	{\simgt}{\lower.5ex\hbox{$\; \buildrel > \over \sim \;$}}
\def\cm{\ifmmode {\rm cm}\else cm\fi}
\newcommand	{\ee}{\ifmmode ^{-2}\else $^{-2}$\fi}
\newcommand{\muesc}{{\mu_{\rm esc}}}
\newcommand{\krho}{{k_\rho}}
\newcommand	{\yr}{\ifmmode {\rm yr}\else y\fi}
\newcommand{\thesc}{{\theta_{\rm esc}}}
\newcommand     {\caln}{\ifmmode {{\cal N}} \else ${{\cal N}}$\fi}
\newcommand{\Rcl}{R_{\rm cl}}
\newcommand	{\e}{\ifmmode ^{-1}\else $^{-1}$\fi}
\newcommand{\rhocl}{{\rho_{\rm cl}}}
\newcommand{\mbarstar}{\overline{m}_\star}
\newcommand{\tAD}{t_{\rm AD}}
\newcommand{\tgstar}{t_{g\star}}
\newcommand{\Avs}{A_{Vs}}
\newcommand{\mstarmax}{m_{\rm \star,\,max}}
\def\rhoo{\rho_{01}}
\newcommand{\rhow}{\rho_{w}}
\newcommand{\pdotw}{\dot{p}_{w}}
\newcommand{\Mdotstar}{\dot{M}_{\star}}
\newcommand{\ecore}     {\varepsilon_{\rm core}}
\newcommand{\lnth}	{\ln(2/\theta_0)}
\newcommand{\Mcore}	{M_{\rm core}}
\newcommand{\mstaro}	{m_{\star,0}}

\newcommand{\Rcore}	{R_{\rm core}}

\begin{document}

 \title{Efficiencies of Low-Mass Star and Star Cluster Formation} 
 \author{Christopher D. Matzner}
\affil{Canadian Institute for Theoretical Astrophysics, University of Toronto}
\and
\author{Christopher F. McKee}
 \affil{Departments of Physics and Astronomy, University of California
 at Berkeley}
 \authoraddr{601 Campbell Hall, U.C. Berkeley, CA 94720}

\begin{abstract} 
Using a quantitative model for bipolar outflows driven by
hydromagnetic protostellar winds, we calculate the efficiency of star
formation assuming that available gas is either converted into stars
or ejected in outflows.  We estimate the efficiency of a single star
formation event in a protostellar core, finding $25\%-70\%$ for cores
with various possible degrees of flattening. The core mass function
and the stellar initial mass function have similar slopes, because the
efficiency is not sensitive to its parameters.  We then consider the
disruption of gas from a dense molecular clump in which a cluster of
young stars is being born. In both cases, we present analytical
formulae for the efficiencies that compare favorably against
observations and, for clusters, against numerical simulations. We
predict efficiencies in the range $30\%-50\%$ for the regions that
form clusters of low-mass stars.  In our model, star formation and gas
dispersal happen concurrently.  We neglect the destructive effects of
massive stars: our results are therefore upper limits to the
efficiency in regions more massive than about $3000~\Msun$.

\end{abstract}

\section{Introduction} \label{S:Intro}
Star formation in molecular clouds is a slow and seemingly inefficient
process.  Clouds are transformed into stars far more slowly than
gravity would allow \citep*[][]{1974ARA&A..12..279Z}; only the deepest
regions of clouds participate in star formation
\citep*[][]{M89,1997ApJ...488..277L};
star-forming regions are disrupted before star formation is complete
\citep*{1984ApJ...285..141L}; and individual protostellar cores may
also shed mass on their way to stardom \citep*{1989ApJS...71...89B,
1998ApJ...495..871L}. 
A growing appreciation for the violence and ubiquity of protostellar
outflows has led many authors to suggest that outflows provide the
mechanism to regulate star formation and to dissipate star-forming
gas, especially in regions that lack massive stars.
Protostellar outflows have been implicated in the support of molecular
clouds and cluster-forming clumps against gravity
\citep*{NS80,LG82,M89,BM96} and in the removal of gas from
star-forming regions 
\citep*{1984ApJ...277..634L,1986ApJ...306L..29L,
1986ApJ...303L..11G,1991ApJ...377..510U,BCD94,BRLB99} or from protostellar
cores \citep*{1988ApJ...324..907M,1995ApJ...450..183N,
1998Natur.392..685V,1998ApJ...495..871L}.
In this paper, we shall construct quantitative models to determine the
destructive effects of outflows on the molecular cores that make
individual stars, and on the larger molecular clumps that make star
clusters.\footnote{The nomenclature for the structure of
molecular clouds has yet to be established.  We adopt
the usage of \cite{WBM}:
{\it Clumps} are coherent regions of molecular gas in
$l-b-v$ space, generally identified from spectral
line maps of molecular emission.  {\it Star-forming clumps}
are the massive clumps out of which stellar clusters form.
Finally, {\it cores} are regions out of which individual stars
(or multiple systems like binaries) form.
We also distinguish protostellar {\it winds}, which are
ejected from the protostar or the protostellar disk,
from protostellar {\it outflows}, which include material
in both the swept-up ambient medium and the wind.}
Disruption by outflows limits the efficiency of low-mass star
formation; it relates the mass function of cores to the stellar
initial mass function; and it determines whether a young stellar
system will remain bound.

\subsection{Disruption of Protostellar Cores} \label{S:Intro:cores}
The most direct progenitors of stars are dense cores within
star-forming regions.  The distribution of initial stellar masses is
therefore controlled by two factors: the distribution of core masses,
and the fraction of a given core that can successfully become a star
\citep*{1995ApJ...450..183N}.  Recent observations of the core mass
distribution by \cite{1998A&A...336..150M} indicate a similarity to
the initial mass function (IMF).

Several authors have made theoretical predictions for the manner in
which a protostar's wind truncates its own accretion, limiting the
core mass that becomes stellar. \cite{1995ApJ...438L..41S} matched the
Kelvin-Helmholtz luminosity of the protostar with the accretion
luminosity of its disk \citep[in the model of][]{1984ApJ...286..529T}
to find the point at which disk accretion stops.  Similarly,
\cite{1996ApJ...464..256A} matched the ram pressure of material
falling directly onto the stellar surface (in the same infall model)
with the wind ram pressure; although it is more physical to consider
ram pressures than luminosities, this theory does not explain what
stops disk accretion. \cite{1995ApJ...450..183N} took a distinctly
different approach: recognizing that material falling onto the disk
might circumvent the outflow, they instead considered the disruption
of the larger-scale density distribution of the accreting core.
Idealizing the core material as static, they argued that accretion
would stop once a momentum-conserving wind bubble (or a stellar HII
region) had traversed its radius.
For fiducial values of the parameters, they find quite
low efficiencies, of order a few percent.

All of these models have considered only spherical protostellar winds,
and they have generally not treated anisotropies in the protostellar
core material (besides centrifugal deflection of the infall). However,
the winds from young stars are collimated into jets on a short length
scale, and the core is likely to be somewhat flattened because of
support from poloidal magnetic fields.  \cite{1996ApJ...472..211L}
incorporated these very elements into a model for the early phase of a
molecular outflow, during which it propagates into its own
protostellar core. However, Li \& Shu did not consider the
implications of this model for the efficiency protostellar accretion.

In \S \ref{S:coredisruption} we shall construct a model for the
disruption of an accreting, somewhat flattened protostellar core.
Assuming that the wind intensity is directly proportional to the
accretion rate, as in the X-wind model \citep{1994ApJ...429..781S},
and in some disk-wind models \citep{1992ApJ...394..117P}, we identify
the angle from the axis within which core material is ejected and
outside of which it is accreted.  As in the model of
\cite{1995ApJ...450..183N}, and also as argued by
\cite{1991ApJ...373..169M}, protostellar accretion is terminated by
the finite size of the protostellar core; however, the efficiency with
which the star forms is a function both of the flattening of the core,
and the ratio of its sound speed to the characteristic wind
velocity. We compare the results of this calculation to observations
of core disruption and constraints on the efficiency of core collapse
in \S \ref{S:coreObs}.

\subsection{Disruption of Cluster-Forming Clumps} \label{S:Intro:clumps}
A protostellar wind can have destructive effects on regions
considerably larger and more massive than its own protostellar core,
repercussions that have implications for the fates of young stellar
clusters.

Stars are born in groups, deeply embedded within dense clumps of
molecular gas. The fate of a nascent stellar system -- whether to
become an expanding association or to remain a bound open cluster --
is determined by how rapidly and effectively it breaks free of the gas
from which it was created. This process is primarily controlled by two
parameters: the efficiency of star formation, defined as the fraction
of gas converted into stars, and the time scale on which the remaining
gas is disrupted.
If the gas removal is rapid relative to the free-fall time $\tff$,
then more than half the mass must be in stars for the cluster to
remain bound \citep[][]{1980ApJ...235..986H}. On the other hand,
sufficiently slow mass loss allows a virialized stellar system to
expand adiabatically and to remain bound.  \cite{1984ApJ...285..141L}
studied intermediate cases, in which they found that part of the
cluster was likely to escape even if a fraction remained bound.

Empirically, roughly only a tenth of all stars remain bound in
clusters \citep{MS78}.  The interactions between stars and gas must
produce this statistic, yet the underlying mechanism by which gas is
dispersed remains largely mysterious.
Most stars are born in large OB associations
\citep[e.g.,][]{1997ApJ...476..144M}, and the violent effects of the
massive stars in these associations have the potential to disrupt the
star-forming gas; however, this can only be in addition to the
disruptive effects of the low-mass stars themselves.

We shall take the embedded clusters in L1640 as prototypical examples
of groups of low-mass stars in the process of formation. As described
by \cite{LEDG91} and \cite{L92}, these clusters have several hundred
stars apiece; they are coincident with the densest and most massive
molecular clumps in the region. Stars currently constitute about
$40\%$ of the total mass in the two largest clumps, but the others
show a much smaller stellar fraction. Moreover, star formation is
restricted to these regions \citep[][]{1997ApJ...488..277L}. This
observation is consistent with the theory of \cite{M89} that stars
form only in regions of high extinction and therefore low ionization,
so that ambipolar diffusion can lead to the creation of protostellar
cores. It lends credence to the idea that the gas and stars can be
treated as a finite, self-gravitating system of a definite mass and
radius \citep[see also][]{1983ApJ...267L..97M},
a notion we shall use in our analysis.

In \S \ref{S:Clumps}, we calculate the amount of gas removed from
a clump by the bipolar molecular outflows driven by the stars
themselves. There is ample evidence for the importance of gas removal
by outflows from star-forming regions: note, for instance, the
disruptions of 
PV Cephei \citep{1984ApJ...277..634L}, B335
\citep{1986ApJ...306L..29L}, Lynds 1221
\citep{1991ApJ...377..510U}, HH83 \citep{BCD94}, and Circinus
\citep{BRLB99}, and the eruptions of the IRS3 outflow from Barnard 5
\citep{1986ApJ...303L..11G} and of the HH300 outflow from Taurus
\citep{AG99}. Indeed, any example of protostellar jets and Herbig-Haro
objects outside the source's parent cloud implies that potentially
star-forming gas has been ejected. Individual protostellar winds are thus
remarkable in their ability to affect pre-stellar cores (on $\sim 0.05~\pc$
scales), cluster forming clumps ($\sim 0.4-1~\pc$) and giant molecular
clouds ($\sim 10-100~\pc$), as evinced by the existence of
parsec-scale flows \citep{1997AJ....114.2708R}.

We shall gauge mass ejection using the model for protostellar outflows
presented in a previous paper \citep{1999ApJ...526L.109M}. This model
generalizes the force distribution predicted in the X-wind model
\citep{1995ApJ...455L.155S} to a much wider class of winds.  The
interaction of the protostellar wind with the ambient medium occurs in
a thin, radiative shell that conserves the wind momentum and ambient
mass in each direction.  In \S \ref{S:mass} we use this model to
calculate the criterion for the outflow shell to be ejected in a
particular direction. The mass loss from such outflows is estimated
analytically in \S \ref{S:Clumps}.  The principal idealization in
our treatment is that each star forms at the exact center of the clump
distribution. Therefore, we present in \S \ref{S:numerics} a numerical
evaluation of the gas lost from a model clump, in which star formation
is distributed according to the local ambipolar diffusion rate
\citep{M89}.  In the Appendix, we evaluate the detailed effect of
gravity on the motion of an outflow shell.

Our analysis suggests that a significant amount of mass can be lost
during the process of star formation itself. Indeed, the rate at which
mass is lost due to outflows is several times the rate at which gas is
converted into stars.  However, we do not consider the disruptive
effects of massive stars, which might cause gas to be removed more
rapidly. Our results are therefore only rough upper limits for the
star formation efficiency in clumps more massive than about
$3000~\Msun$, in which O stars will form.

Unlike most previous work \citep[e.g.,][]{1995ApJ...450..183N}, we
divide the calculation of star formation efficiency into two parts:
the disruption of a star's own protostellar core, and the disruption
of a clump in which an aggregate of young stars is forming. Different
assumptions are appropriate for the two cases. For a core, the rates
of accretion and outflow are directly proportional; the axes along
which the core is flattened and the wind is collimated are likely to
be aligned; the wind duration exceeds the free-fall time; and the
density distribution declines with radius approximately as
$r^{-2}$. For the disruption of a cluster-forming clump, opposite
assumptions apply: for instance, winds are impulsive and not likely to
correlate in direction with the clump's structure. 

\subsection{Estimate of the Star Formation Efficiency}\label{S:EfficEstimate}

Before getting into the details of the calculation, it is worthwhile
to make a simple estimate of the outcome.  The star formation
efficiency for the formation of a single star is
\begin{equation} 
\varepsilon=\frac{\mstar}{\mstar+\Mej},
\label{eq:epsa}
\end{equation}
where $\Mej$ is the mass\footnote{We use lowercase ($\mstar$, $\mw$)
for the mass of an individual star and its wind, and uppercase
($\Mej$, $\Mcl$ and $\Mstar$) for ambient cloud mass or the mass of an
ensemble of stars.}
ejected from the protostellar core or star forming clump due to the
formation of a single star of mass $\mstar$ (see \S \ref{S:efficdefs}
below).  The total momentum in the wind is $p_w=m_wv_w$.  We assume
that the mass of the wind is a fraction $f_w$ of the stellar mass, so
that $m_w=f_w \mstar$ and $p_w=f_w\mstar v_w$.  This momentum is
transferred to the cloud by radiative shocks; as described above, such
a model is in good agreement with observation
\citep{1999ApJ...526L.109M}.  In order for the shocked gas to escape
from the cloud, it must be moving at least at the escape velocity,
$\vesc$.  Since the shocked gas has momentum $p_w$, the ejected mass
is a maximum if all the gas has this minimum velocity, $\Mej\sim
p_w/\vesc.$ Assuming for the moment that this mass is large compared
to the mass of the star that forms, as in the model of
\cite{1995ApJ...450..183N}, we estimate the star formation efficiency
as
\begin{equation}  \label{eq:EpsilonEstimate} 
\varepsilon \simeq\frac{\mstar}{\Mej}\simeq\frac{\mstar c_g\vesc}{p_w}
\simeq\frac{c_g \vesc}{f_wv_w},
\end{equation} 
where the factor $c_g\geq 1$ allows for the decrease in the momentum
of the shocked shell due to the self-gravity of the core or clump.

   This argument can be refined by taking into account the central
fact about protostellar winds, that they are collimated. As originally
shown by \cite{1995ApJ...455L.155S} and demonstrated more generally by
\cite{1999ApJ...526L.109M}, the wind momentum falls off approximately
as $1/(\sin\theta)^2$ for $\theta>\theta_0$, where $\theta$ is
measured from the axis of the wind and $\theta_0\sim 10\ee$ is
discussed in \S \ref{S:Wind-Structure-Duration} below.  The momentum
inside $\theta$ thus scales as $\ln(1/\theta)$: equal amounts of
momentum are contained in each octave of angle.  Since the area
covered by the wind, and therefore the mass swept up, is dominated by
the largest angle, it follows that most of the mass that escapes will
indeed have the minimum possible velocity ($\sim\vesc$), as assumed
above.  Furthermore, since the momentum is spread out logarithmically,
only a fraction $\sim 1/\ln(1/\theta_0)$ of the total momentum is
available to eject matter at $v\sim\vesc$.  As a result, the star
formation efficiency estimated above should be multiplied by this
quantity.  The calculations below show that this fraction is more
precisely given by $\sim 1/\ln(2/\theta_0)$.  Defining the {\it
efficiency factor}~$X$ by
\begin{equation}
X\equiv\frac{c_g\vesc\ln(2/\theta_0)}{f_wv_w},
\label{eq:defX}
\end{equation}
we conclude that for small values of $X$, the star formation
efficiency $\varepsilon\sim X$; at large values of $X$, we expect
$\varepsilon$ to approach its maximum possible value, unity.
These expectations are borne out by the results presented below.
Observe that the collimation of the wind reduces the mass that can be
ejected by a factor $\sim 1/\ln(2/\theta_0)\sim 0.2$ and increases the
star formation efficiency by the same factor.  As a result, our
estimates of $\varepsilon$ are significantly higher than those of
\cite{1995ApJ...450..183N} and are in much closer agreement with
observation.

\section{Instantaneous, Observational and Total Star Formation
Efficiencies}\label{S:efficdefs} 

We begin by discussing the various possible definitions of the star
formation efficiency.  For the collapse of a protostellar core to form
a single star (or a binary)
there is
only one efficiency that can be defined, namely the ratio of the
star's (or binary's) mass to that of its parent core: 
\begin{equation} 
\varepsilon=\ecore \equiv \frac{\mstar}{M_{\rm core}}. 
\end{equation}
This definition is consistent with that in equation (\ref{eq:epsa})
because when the star formation process is complete, all the mass has
been incorporated into the star(s) or ejected from the system.

	For the formation of a stellar cluster in a molecular clump,
we generalize $\varepsilon$ so that it is the {\em instantaneous} star
formation efficiency, which is the rate at which mass is converted
into stars, divided by the rate at which mass is lost from the
gas. (Implicit in this definition is the notion that the star-forming
clump is a well-defined gaseous system.)  We assume that only two
processes change the gas mass: star formation, and mass ejection due
to outflows. Then, the instantaneous star formation efficiency is
\begin{equation}\label{eq:epsilon}
\varepsilon \equiv -\frac{d\Mstar}{d\Mcl} = \frac{d\Mstar}{d\Mstar + d\Mej},
\end{equation}
where $\Mstar$ is the mass of a set of stars, $\Mcl$ is the mass of
the clump, and $\Mej$ is the total mass ejected from the clump by
protostellar winds; we anticipate that 
part, if not all, of the mass
in the winds will be ejected from the clump, and this mass is included
in $\Mej$.  For the formation of a single star, $d\mstar$ becomes
$\mstar$ and $d\Mej$ becomes $\Mej$; as a result, in this case
$\varepsilon$ is the same as that in equation (\ref{eq:epsa}) and is
identical with $\ecore$.  Because of this generality, we shall often
refer to $\Mcl$ as the ``cloud'' mass, using ``cloud'' to refer to
either a clump or a core.  As we shall see below, in our model the
ejected mass is directly proportional to the stellar mass.  As a
result, $\varepsilon$ can be evaluated at any convenient mass, such as
1 $\msun$:
\begin{equation}\label{eq:epsilon2}
\varepsilon= \frac{\msun}{\msun+\Mej(1)},
\end{equation}
where $\Mej(1)$ is the mass ejected from the cloud due to the
formation of a star of 1 $\msun$.  If gas is disrupted by other means,
as might occur where massive stars are forming, then equation
(\ref{eq:epsilon2}) will be an overestimate of $\varepsilon$.

   	The {\em observational} measure of star formation efficiency,
$\SFE$, is the current fraction of the total mass that is stellar:
\begin{equation} \label{eq:SFE} 
\SFE \equiv \frac{\Mstar}{\Mstar+\Mcl}. 
\end{equation} 
This quantity varies from zero to unity as a group of stars forms and
is revealed.  Under the assumption that all the stars that have formed
in the clump can be identified, even if the stars are unbound, the
mass of the clump evolves as $\Mcl = M_{\rm cl,\,0} - \int
\varepsilon^{-1} d\Mstar$, where $M_{\rm cl,\,0}$ is the clump mass
before any stars form.  As a result, we have
\begin{eqnarray} \label{eq:SFE&epsilon} 
\SFE &=& \left(1+\frac{M_{\rm cl,\,0}}{\Mstar} - 
\frac{1}{\Mstar}\int_0^{\Mstar}
\varepsilon^{-1} {d\Mstar'}\right)^{-1}\nonumber\\ 
&\equiv & \left(1+\frac{M_{\rm cl,\,0}}{\Mstar} -
\left<\varepsilon^{-1}\right>_{\Mstar}\right)^{-1}. 
\end{eqnarray} 

The {\em total} efficiency of the star formation process, $\SFEtot$, is
the fraction of the initial clump mass $M_{\rm cl,\,0}$ that will ever
become stars.  Star formation is finished once there is no more gas,
and $\SFE = 1$. From equation (\ref{eq:SFE&epsilon}), we see that this
occurs once $\left<\varepsilon^{-1}\right>_{\Mstar} = M_{\rm
cl,\,0}/\Mstar$; therefore
\begin{equation}\label{eq:SFEtot}
\SFEtot = \left<\varepsilon^{-1}\right>_\Mstar^{-1}, 
\end{equation}
where the average is taken over the entire star formation process. The
instantaneous and total efficiencies are approximately equal, so long
as $\varepsilon$ is relatively constant.

On the other hand, $\varepsilon$ can be considered a typical
value for the observed $\SFE$ only if both are small. For instance,
consider a clump in which $\varepsilon$ is constant and star formation
is half complete ($\Mstar = \SFEtot M_{\rm cl,\,0}/2 = \varepsilon
M_{\rm cl,\,0}/2$). Equation (\ref{eq:SFE&epsilon}) shows that $\SFE =
1/(1+\varepsilon^{-1})$ at this point.

\section{Structure, Intensity and Duration of protostellar winds}
\label{S:Wind-Structure-Duration} 
The angular distribution of the momentum $\pw$ in collimated
protostellar winds can be written as
\begin{equation}\label{eq:dpwdOmega}
\frac{d\pw}{d\Omega} = \frac{\pw(t)}{4\pi} P(\mu),
\end{equation}
where $\mu\equiv\cos\theta$ labels directions relative to the wind
axis, and $P(\mu)$ is a force distribution normalized such that
$\int_0^1 P(\mu)d\mu = 1$.  \cite{1999ApJ...526L.109M} and
\cite{myphd} showed that any hydromagnetic protostellar wind should
limit to a common force distribution on scales large compared to the
source:
\begin{equation}\label{eq:PmuHydromag}
P(\mu) = \frac{1}{\ln({2}/{\theta_0}) (1+\theta_0^2-\mu^2)}, 
\end{equation} 
with $\theta_0\ll 1$.  Moreover, under the assumption of a locally
momentum conserving interaction with the ambient cloud
\citep{Shuea91}, this relation was shown to give good agreement with
the observed position-velocity and mass-velocity distributions of
bipolar molecular outflows.
In cases where we are interested only in angles away 
from the axis ($1-\mu^2\gg
\theta_0^2$), such as the disruption of a protostellar core by its wind
(\S \ref{S:coredisruption}), equation (\ref{eq:PmuHydromag}) simplifies
to, 
\begin{equation}\label{eq:PmuAppxn}
P(\mu)\simeq \frac{1}{\ln({2}/{\theta_0}) (\sin\theta)^2}. 
\end{equation}
The dependence $\pw\propto(\sin\theta)^{-2}$ is caused by the balance
of magnetic stresses in the wind at large distances from its source,
as in the model of \cite{1995ApJ...455L.155S}.  Deviations from this
power-law relationship can be caused by variations of the conserved
variables across streamlines (e.g., because of differences in initial
disk radius), but these deviations are typically quite small. The
broadening angle $\theta_0$ could be the result of a wandering of the
jet, internal shocks from variations in the wind velocity, or
hydromagnetic instabilities; \cite{1999ApJ...526L.109M} estimated
$\theta_0\simeq 10^{-2}$ on the basis of comparisons with observations.

The wind momentum $\pw$ is the product of the wind velocity $\vw
\simeq 200 ~\kms$ and the wind mass $\mw$, which itself is a fraction
$\fw$ of the final mass $\mstar$ of the protostar that drives the
wind. Therefore the net wind momentum is
\begin{equation}\label{eq:pwind}
\pw = \fw \vw \mstar. 
\end{equation}
We shall assume for simplicity that $\fw\vw$ does not depend on
$\mstar$.
This assumption is justified in the X-wind model
\citep{1994ApJ...429..781S} by the insensitivity of both $f_w$ and
$v_w$ to the details of the accretion.  In this model, $f_w$ is set by
the geometry of field lines in a very small region around the disk
truncation radius. Also, deuterium burning introduces a thermostat
that maintains the escape velocity from the stellar surface near a
constant value \citep{1988ApJ...332..804S}, and the ratio between
$v_w$ and the escape velocity is insensitive to the star's mass,
magnetization and accretion rate.  \cite{MBM2000} argue that the model
of \cite{1994ApJ...429..808N} implies $\fw\vw\simeq 40~\kms$ if one
takes the average wind luminosity to be half its final value.  For a
wind velocity of $200~\kms$, this implies $\fw\simeq 1/5$, between the
predictions $\fw \simeq 1/3$ and $\fw \simeq 1/10$ of the original
X-wind \citep{1988ApJ...328L..19S} and disk wind
\citep[e.g.,][]{1992ApJ...394..117P} theories, respectively.

The wind momentum output may vary rapidly, but it declines in
intensity over a wind or accretion time scale $\tw$, which is generally
taken to be about $10^5$ years. There is little observational evidence
that $\tw$ varies with $\mstar$; \cite{1996A&A...311..858B} argue for
an exponential decline in outflow intensity, on a time scale of
$0.9\times 10^5 ~\yr$, throughout their sample of Class 0 and I
protostars.  Thus, for star-forming clumps less dense than $10^5
~\cm^{-3}$, whose free-fall times are longer than $10^5$ years, it is
clearly a good approximation to consider the deposition of an
outflow's momentum as impulsive ($\tw<\tff$).

What happens at higher densities? The duration of an intense
protostellar wind is identical to the time scale of accretion onto the
protostar. This in turn is essentially identical to the duration of
accretion onto the combined star-disk system, for a stable accretion
disk cannot hold most of the total mass
\citep{1989ApJ...347..959A,1990ApJ...358..495S}.  Finally, this
accretion time in a magnetized core is roughly twice the thermal
crossing time of the protostellar core if ambipolar diffusion is not
considered \citep{1997ApJ...475..237L}, and less if it is
\citep{1997ApJ...485..660S, 1998ApJ...497..850L}; thus, the accretion
and dynamical times in a protostellar core are similar. But, the
dynamical time of a protostellar core is shorter than that of a
molecular clump if the former is an overdense region within the
latter. Therefore, we shall assume that protostellar winds are always
impulsive events compared to the dynamical time of the parent
star-forming clump, even for mean clump densities above $10^5 ~
\cm^{-3}$.  This assumption, which remains to be tested by
observations, will simplify our analysis of the efficiency of stellar
cluster formation in \S \ref{S:Clumps}.

\section{Mass Ejection by Protostellar Outflows}
\label{S:mass}

To estimate the amount of mass ejected from either a protostellar core
or a star-forming clump we approximate the shocked wind and cloud
material as a thin shell whose motion is purely radial.  In other
words, neither mass nor momentum is mixed between angular sectors, and
there is no lateral momentum generated in the shell.  These
approximations were introduced by \cite{Shuea91}, and were shown by
\cite{1999ApJ...526L.109M} to reproduce the observational features of
molecular outflows.  For the case in which the ambient gas is not
isotropic, we shall consider only the monopole component of gravity;
\cite{1997ApJ...475..237L} demonstrated that this is a viable
approximation, even in the extreme case of a disk-like core.

Our approach is as follows.  A fixed fraction $f_w$ of the accreted
material is ejected at velocity $v_w$ in a wind with angular
dependence given by equation (\ref{eq:PmuHydromag}). Assuming thin,
radiative shocks, purely radial motion and monopole gravity, we shall
identify the direction (characterized by the direction cosine
$\muesc\equiv\cos\thesc$) in which the wind is marginally able to
drive the ambient material outward at the escape velocity of the
cloud. Only material further toward the equator can be accreted, since
the rest is ejected. And, only this accreted material is available to
be reprocessed in the wind. The fraction of core material that
successfully accretes will thus be determined self-consistently.

To simplify the discussion of the condition for a protostellar wind to
eject gas from the cloud, we shall defer including the effect of
gravity on the propagation of the shock in the cloud to the {\it
Appendix}.  In the absence of the perturbation introduced by gravity,
the shell's velocity $v_s$ is set by the total emitted wind momentum
and the ambient mass swept up in each direction in the direction
specified by $\mu$:
\begin{equation}\label{eq:vsNoGrav}
\frac{dM_{\rm cl}}{d\Omega}v_s(r,\mu,t) = \frac{d\pw(t)}{d\Omega}.
\end{equation}
Here we have assumed that the mass and flight time of the wind can be
neglected in comparison with the ambient mass and age of the
outflow. This is a good assumption in directions in which
$v_s\sim\vesc$, since the mass of the wind must be negligible if it is
to drive a shock at a velocity small compared to the wind velocity
\citep{1999ApJ...526L.109M}. We have also neglected the effects of the
ambient pressure or magnetic field, which should be a reasonably good
approximation so long as the shock is traveling faster than the local
signal speed.  Note that these perturbations are similar in magnitude
to gravity if the clump is in virial equilibrium.

	We express the mass distribution of the ambient gas as
\begin{equation}
\frac{dM_{\rm cl}}{d\Omega}=\frac{1}{4\pi} Q(\mu)\Mcl(r),
\label{eq:dmdo}
\end{equation}
where $Q$ is normalized by $\int_0^1Q(\mu) d\mu=1$ and $M(r)$ is the
total mass inside $r$.  Since $dp_w/d\Omega\propto P(\mu)$
(eq. \ref{eq:dpwdOmega}), the equation of motion for the shell becomes
\begin{equation}
Q(\mu)\Mcl v_s=P(\mu)p_w(t).
\label{eq:QMv}
\end{equation}

	The condition for a sector of the shocked gas to escape from
the gravitational field of the cloud in which the protostar is
embedded can be estimated by setting $p_w(t)$ equal to the total
momentum of the wind, $m_wv_w=\mstar f_w v_w$, and by setting the
shock velocity equal to the escape velocity, $\vesc$.  The effects of
gravity on the propagation of the shock are discussed in the {\it
Appendix}; they can be incorporated by including a factor $c_g\geq 1$
that is evaluated there:
\begin{equation}
\frac{P(\muesc)}{Q(\muesc)}\mstar f_wv_w=c_g\Mcl\vesc.
\label{eq:esc}
\end{equation}
For impulsive winds (those that last a time that is short compared to
the free-fall time of the ambient cloud, $t_w\ll\tff$), the factor
$c_g$ is close to unity.  For steady winds ($t_w\gg \tff$), $c_g$ is
proportional to the ratio $t_w/\tff$: only the momentum transferred
during a free-fall time is effective in ejecting cloud material.

In deriving this escape condition, we have assumed that the cloud
material does not have a significant infall velocity in the region of
interest ($\theta < \thesc$). In the theory of
\cite{1977ApJ...214..488S}, the core material is completely at rest
outside an expansion wave moving at the thermal velocity, $a$. Since
$\vesc > a$, the outflow shell encounters no infall whatsoever in this
direction.  However, more complete theories
\citep{1997ApJ...485..660S,1998ApJ...497..850L} and recent
observations \citep{1999ApJ...526..788L} suggest an inflow velocity
intermediate between the ambipolar drift velocity and the isothermal
sound speed. This is still significantly slower than $\vesc$,
justifying our approximation that material upstream of the outflow
shell is motionless in the direction of interest.

To evaluate the escape condition (eq. \ref{eq:esc}), we introduce the
efficiency parameter $X$ defined in equation (\ref{eq:defX}) and use
the expression for $P(\mu)$ in equation (\ref{eq:PmuHydromag}):
\begin{equation}
(1+\theta_0^2-\muesc^2)Q(\muesc)=\frac{\mstar}{X\Mcl}.
\label{eq:muesca}
\end{equation}
Numerically, we have
\begin{equation}
X=0.132 c_g\,\frac{\lnth}{\ln 200}\left(\frac{\vesc}{1\ \kms}\right)
	\left(\frac{40\ \kms}{f_wv_w}\right).
\label{eq:Xnum}
\end{equation}

The mass ejected from the cloud by the protostellar 
wind is the sum of the swept-up cloud mass inside $\thesc$ and
the wind mass inside $\thesc$:
\begin{equation}
\Mej=\Mcl\int_\muesc^1 Q(\mu) d\mu + (1-\phi_w) m_w,
\end{equation}
where
\begin{equation}
\phi_w\equiv\int_0^\muesc P(\mu)d\mu
\end{equation}
is the fraction of the wind momentum retained in the cloud; we assume
that $v_w$ is independent of $\mu$ and also that the shocked wind is
dynamically cold, in which case $\phi_w$ is also the fraction of the
wind mass that is retained.

The value of the ejected mass determines the instantaneous star
formation efficiency (eq. \ref{eq:epsilon}),
\begin{eqnarray}
\frac{1}{\varepsilon}&=&1+\frac{\Mej}{\mstar} \\
	&=& 1+\frac{\Mcl}{\mstar}\int_\muesc^1 Q(\mu) d\mu +
	(1-\phi_w)f_w. 
\end{eqnarray}
Inserting equation (\ref{eq:muesca}), we find
\begin{eqnarray}
\frac{1}{\varepsilon}&=& 1+\frac{1}{X(1+\theta_0^2-\muesc^2)Q(\muesc)}
	\int_\muesc^1 Q(\mu) d\mu \nonumber \\ &~& + (1-\phi_w)f_w.
\label{eq:epsilonb}
\end{eqnarray}

If the cloud is isotropic [$Q(\mu)=1$], the ejected mass is simply
$\Mej=\Mcl(1-\muesc) + (1-\phi_w)m_w$.  Neither cloud nor wind mass is
ejected if $\muesc=1$, since $\phi_w=1$ if the outflow does not reach
the surface of the cloud; according to equation (\ref{eq:muesca}),
this occurs at a stellar mass
\begin{equation}
\mstaro=X\theta_0^2\Mcl.
\label{eq:mstaro}
\end{equation}
For $X\simlt 1$, $\theta_0\simlt 10\ee$, and $\Mcl\simlt 10^3\;\msun$,
it follows that $\mstaro\simlt 0.1\;\msun$, so that most stars are
more massive than $\mstaro$.  Equation (\ref{eq:muesca}) implies that
the mass ejected from an isotropic cloud is
\begin{equation}
\Mej=\frac{\mstar-\mstaro}{X(1+\muesc)} + (1-\phi_w)f_w m_\star
\label{eq:Mej}
\end{equation}
for $\mstar\geq\mstaro$.  
Note that if
$\mstar\gg\mstaro$, then the amount of
mass ejected from the cloud is directly proportional to the amount of
mass converted into stars. The resulting star formation efficiency for
an isotropic cloud is
\begin{equation}
\frac{1}{\varepsilon}\simeq 1+\frac{1}{X(1+\muesc)}
\left(1-\frac{\mstaro}{\mstar}\right)+(1-\phi_w)f_w
\label{eq:epsilonc}
\end{equation}
for $\mstar\geq\mstaro$; for $\mstar\leq\mstaro$, we have $\varepsilon=1$.
Weakening the wind or increasing the escape velocity (i.e., increasing
$X$) increases the stellar mass $\mstaro$ that cannot eject any
material.  For a star with $\mstar>\mstaro$, it causes less material
to be ejected: hence $\muesc$ increases with $X$ (as in equation
[\ref{eq:OneOverMuesc}]). This also reduces the amount of wind mass
lost: therefore, $\phi_w$ increases with $X$. As a result, we see that
the efficiency $\varepsilon$ is an increasing function of $X$.

\section{Core Disruption and the Efficiency of Single Star Formation}
\label{S:coredisruption} 

We now consider the disruption of a protostellar core by a single
protostar.  The results of the previous section remain valid, but we
now have an additional condition: all the mass in the cloud with
$\theta>\thesc$ either goes into the star or 
into the wind that escapes:
\begin{equation}
\mstar+m_w(1-\phi_w)=\Mcore\int_0^\muesc Q(\mu)d\mu.
\label{eq:msmw}
\end{equation}
For a protostellar core, $\thesc$ is not small, as we shall see below.
Since the wind is strongly concentrated toward the axis according to
equation (\ref{eq:PmuHydromag}), it follows that the fraction of the
wind material that is retained by the core is generally small,
$\phi_w\ll 1$.  We therefore neglect $\phi_w$ and find the star
formation efficiency of the core to be ($\ecore=\varepsilon$):
\begin{equation}
\ecore\simeq\frac{1}{(1+f_w)}\int_0^\muesc Q(\mu) d\mu.
\label{eq:ecore}
\end{equation}

\subsection{Isotropic Protostellar Cores}

	For the case of an isotropic core, equation (\ref{eq:msmw})
reduces to $\mstar(1+f_w)=\Mcore\muesc$, where we have set $\phi_w=0$
as discussed just above.  The mass of a protostar in a core is always
much greater than $\mstaro$: By definition, $\mstar$ could 
be of order $\mstaro$ only if $\muesc\simeq 1$; but then $\mstar\simeq
\Mcore$ since $f_w<1$, so that $\mstar/\mstaro\simeq 1/X\theta_0^2\gg
1$, a contradiction.
Since $\ecore=\varepsilon$, we can use equations (\ref{eq:epsilonc})
(neglecting $\mstaro$) and (\ref{eq:ecore}) to obtain an equation for
$\muesc$:
\begin{equation}\label{eq:OneOverMuesc}
\frac{1}{\muesc} = 1 + \frac{1}{X(1+f_w)(1+\muesc)} ; 
\end{equation} 
solving for $\muesc$ and $\ecore$, 
\begin{equation} \label{eq:ecoreb}
\ecore = \frac{\muesc}{1+f_w} = \frac{2 X}{1 + [1+4(1+f_w)^2 X^2]^{1/2}}
\end{equation}  
Thus, the efficiency of star formation in an isotropic core that forms a
single star is always less than the efficiency parameter, $\ecore\leq
X$; moreover, it is always less than $1/(1+f_w)$ since the wind mass
is easily ejected.

\subsection{Self-Similar Magnetized, Isothermal Cores}\label{SS:LScores}

	Equation (\ref{eq:ecoreb}) provides a general expression for
the efficiency of star formation in an isotropic protostellar core.
It depends on the properties of the core through the parameter
$X\propto c_g\vesc$.  To determine numerical values for the star
formation efficiency, and to generalize to non-isotropic cores, we
must adopt a specific model for the core.  A class of models that is
particularly suitable for this purpose is the class of magnetized,
isothermal core distributions considered by
\cite{1996ApJ...472..211L}.  In these models, magnetic support causes
the density distribution to be flattened relative to the singular
isothermal sphere, and it increases the mean density by a factor
$(1+H_0)$:
\begin{equation}\label{eq:flattenedcore}
\rho(r,\mu) = (1+H_0)Q(\mu) \frac{a^2}{2\pi G r^2}, 
\end{equation}
where $a$ is the isothermal sound speed (or effective sound speed in
the case of turbulent support).
\cite{1996ApJ...472..211L} have calculated $Q(\mu)$ and $(1+H_0)$ for
various values of the dimensionless mass-to-flux ratio. The results
of these calculations can be approximated by
\begin{equation} \label{eq:Qappxn} 
Q\simeq \frac{(\sin \theta)^{k_\theta}}{\int_0^{\pi/2}
(\sin\theta)^{k_\theta+1} d\theta}, 
\end{equation} 
where
\begin{equation} \label{eq:kthetafit}
k_\theta = 9.4 H_0^{5/4}. 
\end{equation}
This fit was made for the values of $H_0$ presented by
\cite{1996ApJ...472..211L}, and is therefore strictly valid only for
$H_0\leq 1$. It was chosen to make the approximation of equation
(\ref{eq:Qappxn}) fit the correct $Q(\theta)$ best near the equator for
each value of $H_0$, because we expect the boundary between accreted
and ejected material to be near the equator. 
This fit is accurate to within $15\%$ wherever it predicts
$Q(\theta)>0.25$; for all values $0<H_0<1$, this condition is
satisfied for $\theta>45^\circ$. Close to the equator, this fit is good
to within several percent -- except that when $H_0=1$, it is $8\%$ below
the \citeauthor{1996ApJ...472..211L} value. 

The mass enclosed within a radius $r$ of a Li-Shu core is 
\begin{equation} \label{eq:coreM}
\Mcore(r) = 2(1+H_0) \frac{a^2}{G} r.
\end{equation}
\cite{1997ApJ...475..237L} have argued that the mass accretion rate in
the self-similar collapse of such a core is always within $5\%$ of the
value
\begin{equation}\label{eq:MdotappxNoEjection}
\dot{m}_{\rm acc} \simeq (1+H_0) \frac{a^3}{G}.
\end{equation}
Comparing the accretion rate to the mass in the core, we see that the
star-disk system swallows all of the mass initially within a radius
$r=at/2$ at a time $t$ after the onset of accretion.  Equation
(\ref{eq:MdotappxNoEjection}) should properly be considered a lower
limit to the accretion rate, as it assumes perfect coupling between
mass and flux; ambipolar diffusion would allow for faster accretion
\citep{1997ApJ...485..660S,1998ApJ...497..850L}.  However, as long as
the core is not highly magnetized \citep[$H_0 \lesssim 1$, as expected
from observations; ][]{1997ApJ...475..237L}, the idealization of
strong coupling is a good one \citep{1998ApJ...497..850L}.

Equation (\ref{eq:MdotappxNoEjection}) assumes that mass accretion
occurs in all directions.  However, material within an angle $\thesc$
of the axis is ejected rather than accreted. The gas outside $\muesc$
is the sum of the star mass and the wind mass (eq. \ref{eq:msmw}), so
that
\begin{equation}\label{eq:varepsiloncDefn}
\ecore(1+f_w) = \int_0^{\muesc} Q(\mu)d\mu 
	= \frac{\int_{\theta_{\rm esc}}^{\pi/2}
	(\sin\theta)^{k_\theta+1} d\theta}
	{\int_0^{\pi/2} (\sin\theta)^{k_\theta+1} d\theta}.
\end{equation} 
(Note that we have set $\phi_w=0$ in this equation.)
Accounting for this division, the mass accretion rate is somewhat
smaller than cited above:
\begin{equation} \label{eq:Mdotappx}
\dot{m}_{\rm acc} = \ecore(1+f_w) (1+H_0) \frac{a^3}{G}. 
\end{equation} 
This expression is based on the simplifying assumption that the wind
does not slow accretion in any direction that it does not eject
material.  Since the wind mass is related to the star mass by $m_w =
f_w \mstar$ and since the two share the total accreted mass ($m_{\rm
acc} = m_w+\mstar$), we have $\dot \mstar=m_{\rm acc}/(1+f_w)$ and
$\dot{m}_w = \dot{m}_{\rm acc} f_w/(f_w+1)$.

To determine the star formation efficiency, we must evaluate the
efficiency parameter $X$ for the Li-Shu cores.  We first determine the
factor $c_g$ that incorporates the effects of gravity on the
propagation of the wind-driven shock.  The wind is impulsive or steady
depending on the value of $t_w/\tff$.  As remarked above, mass inside
a radius $r$ is accreted at a time $2r/a$; under the assumption that
the core is terminated at a radius $\Rcore$, the wind lifetime is then
\begin{equation}
t_w=\frac{2\Rcore}{a}.
\end{equation}
By comparison, the free-fall time 
in the monopole approximation is
\begin{equation}
\tff=\left(\frac{3\pi}{32 G\bar\rho}\right)^{1/2}
	=\frac{\pi}{4(1+H_0)^{1/2}}\;
	\left(\frac{\Rcore}{a}\right).
\end{equation}
Since $t_w$ is several times $\tff$, the wind may be considered
steady.  We make the conservative assumption that the wind decouples
from the shocked shell once it leaves the core, so that $c_g$ is given
by equation (\ref{eq:shellsteadyEscape2}).  Because the Li-Shu core
has $\rho\propto r^{-2}$ (i.e., $\krho=2$), this becomes
\begin{equation}
c_g=\left(\frac{3\pi}{4}\right)\frac{t_w}{\tff}
	=6(1+H_0)^{1/2}.
\end{equation}
Noting that $\vesc/a=2(1+H_0)^{1/2}$, we obtain
\begin{eqnarray}\label{eq:defXb}
{ X} &=& 12 (1+H_0)\ln(2/\theta_0) \frac{a}{\fw\vw}
\nonumber\\ 
&=& 0.32(1+H_0)~
\frac{\ln(2/\theta_0)}{\ln(200)}~\left(\frac{a}{0.2~\kms}\right)\nonumber \\ 
&~&~~~~~~~~~~~\times
~\left(\frac{40~\kms}{\fw\vw}\right).
\end{eqnarray}

\begin{figure*}
\centerline{\epsfig{figure=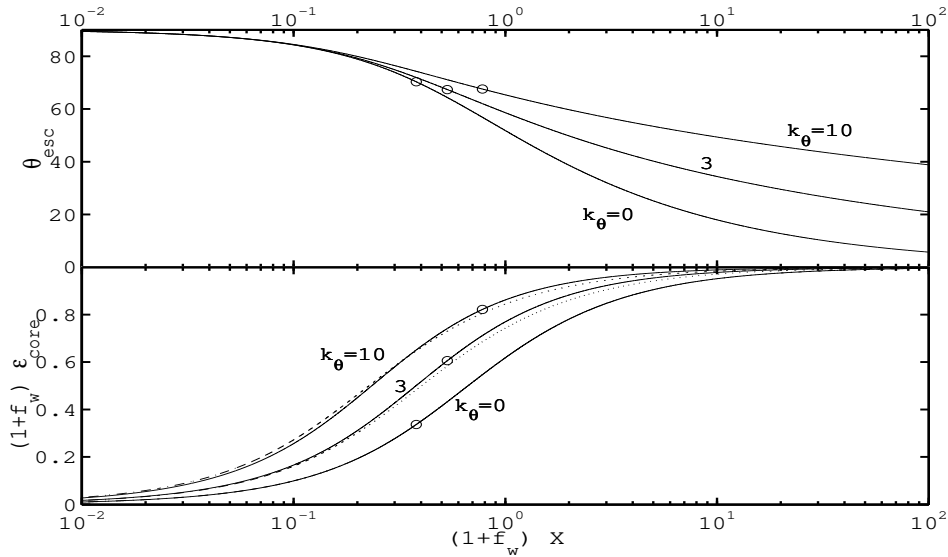,width=5in, height = 3in, angle=0}}
\caption[Efficiency of single star formation in flattened cores] {Core
disruption and the efficiency of single star formation. {\em Top}: the
angle $\theta_{\rm esc}$ (equation [\ref{eq:CoreThetaEsc}]) dividing
accreted and ejected core material, which corresponds to an outflow
velocity equal to the escape velocity of the collapsing core.  For
various values of the core flattening parameter $k_\theta$, the
fiducial value $X=0.32(1+H_0)$ (equation [\ref{eq:defXb}]; {\em
circles}) leads to $\theta_{\rm esc} \simeq 67^\circ$ from the
axis. {\em Bottom}: The efficiency $\ecore$ of core collapse
(equations [\ref{eq:varepsiloncDefn}] and [\ref{eq:CoreThetaEsc}];
{\em solid lines}) and an approximate formula (equation
[\ref{eq:epsiloncAppc}], {\em dotted lines}).  Curves are plotted for
$0<k_\theta<10$, corresponding to $1<(1+H_0)<2$, where approximation
(\ref{eq:Qappxn}) is known to be accurate.
\label{fig:coreeffic} }
\end{figure*}

The star formation efficiency (eq. \ref{eq:varepsiloncDefn}) and the
escape condition (eq. \ref{eq:esc}) together determine the escape
angle $\theta_{\rm esc}$:
\begin{equation}\label{eq:CoreThetaEsc}
\frac{\int_{\theta_{\rm esc}}^{\pi/2}
(\sin\theta)^{k_\theta+1}d\theta}{(\sin\thesc)^{k_\theta+2}} = 
	(1+f_w){X}.
\end{equation}
We find that the solution to this equation gives a star formation
efficiency of the form in equation (\ref{eq:ecoreb}) with the
replacement $X\rightarrow (1+H_0)^{3/2} X$:
\begin{equation}\label{eq:epsiloncAppc}
\ecore \simeq \frac{2(1+H_0)^{3/2}X}{1+[1+4(1+H_0)^3 (1+f_w)^2 X^2]^{1/2}}.
\end{equation}
The validity of this approximation, depicted in Figure
\ref{fig:coreeffic}, implies that an increase in mean density by a
factor of $(1+H_0)$ is equivalent, as far as the efficiency is
concerned, to a weakening of the momentum output in the wind by the
factor $(1+H_0)^{5/2}$.  One of these factors of $(1+H_0)$ 
is contained in $X$; it arises from the increase in escape velocity
compared to a fixed value of $\vw$, as the mean density is
increased. The other one and a half powers of $(1+H_0)$ arise because
of the increasing concentration of infalling material toward the
equator, in the direction the wind is weakest.
In the limit of a completely disk-like core [$(1+H_0)\gg 1$], outflow
occurs in all directions but along the equator ($\theta_{\rm
esc}\rightarrow \pi/2$), but only the wind mass is ejected [$\ecore
\rightarrow 1/(1+f_w)$]. 

\subsection{The Initial Mass Function} \label{S:IMF}
\newcommand{\nstar}	{\caln_*}
\newcommand{\ncore}	{\caln_{\rm core}}
\newcommand{\acore}	{{\alpha_{\rm core}}}
\newcommand{\astar}	{{\alpha_{*}}}
\newcommand{\eprime}	{{\varepsilon '}}

Having determined $\ecore$, it is possible to relate the IMF to the
mass distribution of protostellar cores \citep{1995ApJ...450..183N}.
Let $d\nstar$ be the number of stars formed with masses in the mass
range $\mstar$ to $\mstar+d\mstar$.  Using a similar notation for the
protostellar cores, we assume that they have a power-law distribution:
\begin{equation}
\frac{d\ncore}{d\ln\Mcore}\propto \Mcore^{-\acore}.
\end{equation}
Let
\begin{equation}
\eprime\equiv\frac{d\ln\ecore}{d\ln\Mcore}.
\end{equation}
If we approximate $\eprime$ as constant, then
$\ecore
=m_\star/\Mcore
\propto\Mcore^\eprime$.  The IMF is then
\begin{equation}
\frac{d\nstar}{d\ln \mstar}=\frac{d\ncore}{d\ln \mstar}
	\propto \mstar^{-\acore/(1+\eprime)}.
\end{equation}

	To evaluate $\eprime$, we assume that the flattening parameter
$H_0$ is independent of the core mass.  Then $\ecore$ depends on $X$,
and it is straightforward to show from equation
(\ref{eq:epsiloncAppc}) that $d\ln \ecore/d\ln X \leq 1$, and
additionally that $d\ln \ecore/d\ln X$ is a decreasing function of
both $X$ and $(1+H_0)$.  From the definition (eq. \ref{eq:defX}), we
see that $X$ depends directly on the properties of the core only
through $\vesc$.  In principle, there could be an implicit dependence
through $f_wv_w$, but we have argued in \S
\ref{S:Wind-Structure-Duration} that this is a constant for low-mass
stars.  Since the protostellar cores are gravitationally bound, their
internal velocity dispersion is a fraction of their escape velocity,
$\sigma\propto\vesc$.  Here, $\sigma$ includes both thermal and
turbulent motions. Since protostellar cores are never colder than 10
K, $\sigma \geq 0.2~\kms$: clumps must lie to the right of the circles
in figure \ref{fig:coreeffic}.  Therefore, $\ecore$ cannot be less
than $0.3$ and $d\ln\ecore/d\ln X$ cannot exceed $0.75$, the values
for an unflattened core with $\sigma = 0.2~\kms$.  Now, protostellar
cores obey a line-width size relation of the form $\sigma\propto
\Rcore^p$ \citep{1981MNRAS.194..809L}.  Since $\vesc\propto
(\Mcore/\Rcore)^{1/2}$, we have $X\propto \vesc\propto
\Mcore^{p/(2p+1)}$, so that
\begin{equation}
\eprime=\frac{d\ln\ecore}{d\ln X}\;\frac{d\ln X}{d\ln \Mcore}
	\leq 0.75 \frac{d\ln X}{d\ln \Mcore}=0.75 \frac{p}{2p+1}.
\end{equation}
If the motions are primarily thermal, as we assumed in \S
\ref{SS:LScores}, then $p\simeq 0$ and therefore $\eprime\simeq 0$.
On the other hand, if the motions are primarily non-thermal, then
$p\simeq 1/2$ \citep{1995ApJ...446..665C} and $\eprime\leq 0.19$.  As
a result, for all values of $\ecore$, and regardless of whether the
line width is primarily thermal or nonthermal, we have
\begin{equation}\label{eq:EprimeForSmallEcore}
0\leq\eprime\leq 0.19.
\end{equation}
so that ${d\caln_*}/{d\ln \mstar}\propto \mstar^{-(0.84-1)\acore}$.
Since in practice the thermal contribution to the line width is
significant for low-mass cores, and since $d\ln\ecore/d\ln X$
decreases as $\sigma$ (and thus $X$) increases, which reduces
$\eprime$ for higher mass cores, we expect $\eprime\la 0.1$ in
general. Then, 
\begin{equation}
\label{eq:IMF}
\frac{d\caln_*}{d\ln \mstar}\propto \mstar^{-(0.9-1)\acore}.
\end{equation}
Thus, over the mass range in which the core mass function is a power
law and the star formation efficiency is determined by outflows, the
slope of the IMF is expected quite generally to be just slightly
flatter than than that of the protostellar cores. 
This theory does not address the IMF of high-mass stars, nor of cores
sufficiently turbulent for fragmentation to occur during their collapse. 

Although the above argument assumes a constant value of $H_0$ among
cores, the result that the IMF is very close in slope to the core mass
function is independent of this assumption. This follows from the fact
that $0.3\leq \ecore < 1$ (figure \ref{fig:coreeffic}). The two slopes
cannot differ greatly if the stellar mass is always within a factor of
three of the core mass. 

Features in the core mass function (such as a peak) appear in the IMF
at a mass that is smaller by the factor $\ecore$.

\subsection{Comparison with Observations} \label{S:coreObs}
We have adopted a scenario in which a protostellar core is destroyed
simultaneously by accretion onto a central star (near the equator) and
by protostellar outflow (near the poles). This is broadly consistent
with observations that indicate both processes are important in
dissipating cores, such as those of 
\cite{1988ApJ...324..907M} and
\cite*{1998ApJ...495..871L}. This geometry is confirmed by
observations of the IRS 2 outflow source in Barnard 5
\citep{1998Natur.392..685V}, which clearly discern an equatorial inflow
coexisting with an axial outflow.

Although we have not calculated the outflow structure in detail, the
current theory implies the existence of a massive, slow, conical
outflow that reaches the core escape velocity at about $23^\circ$ from
the equator for the fiducial values of the parameters.  This
structure, which also appears in the theory of
\cite{1996ApJ...472..211L}, is likely to correspond to the massive,
poorly collimated, $1~\kms$ biconical outflow observed around T Tauri
by \cite{1996ApJ...470.1001M}. The outflow cone observed by Velusamy
and Langer extends to $\sim 27^\circ$ from the equator, in rough
agreement with our prediction. More specific comparisons would require
a solution to the problem of simultaneous infall and outflow, which is
beyond the scope of this paper. 

We derive values of $\ecore$ that vary from $25\%$ to $75\%$ for the
fiducial value $X=0.32(1+H_0)$, for values of the core flattening
parameter $0< k_\theta<10$ (i.e., $0<H_0<1$). Thus, we expect that
spherical cores will lose three-quarters of their mass while forming a
star, whereas cores at the upper end of the acceptable range of
flattening \citep{1996ApJ...472..211L} will lose only a quarter of
their mass. The stellar initial mass function therefore depends not
only on the mass function for cores, but also on the relative degrees
of flattening of low and high-mass cores, and on changes of cores'
line widths with mass. Observations that indicate a strong similarity
between the core and star mass functions (in both shape and peak
mass), such as those of \cite{1998A&A...336..150M}, require that
$\ecore$ be roughly constant and of order unity. This is certainly
consistent with the above theory, since $X(1+H_0)^{3/2}$ is typically
of order unity and $f_w$ is small, so that $\ecore$ is indeed close to
unity (eq. \ref{eq:epsiloncAppc}).  Observations that indicate only
similar slopes between the two distributions
\citep[e.g.,][]{1998ApJ...508L..91T} do not constrain the value of
$\ecore$.  They only require that it does not vary rapidly, a result
that is consistent with the results of \S \ref{S:IMF}.

\section{Instantaneous Star Formation Efficiency in Cluster-Forming Clumps}
\label{S:Clumps}  

Clumps massive enough to form clusters of stars differ from the
protostellar cores considered in the last section in several key
respects.  First, it is unlikely that there is a strong correlation
between the geometry of the clump and the direction of the outflow.
Provided the clump is not highly flattened, we can obtain a reasonable
approximation to the effect of the wind by assuming that the clump is
spherical [$Q(\mu)=1$].  As a result, the star formation efficiency is
given by equation (\ref{eq:epsilonc}).  Second, much of the mass of
the clump is generally not involved in the formation of the star.  An
upper limit on the mass of a star that forms in the clump is given by
the result for a protostellar core, in which all the mass is either
accreted or ejected:
\begin{equation}
\mstar\leq\mstarmax\equiv\ecore \Mcl=\frac{2X\Mcl}{1+[1+4X^2(1+f_w)^2]^{1/2}}.
\label{eq:mstarmax}
\end{equation}
There is no restriction on the formation of stars less massive than
this, but stars with $\mstar<\mstaro$ have outflows too weak to eject
any matter from the clump.

A further difference between clumps and cores is that since the clumps
contain the cores, the cores are generally significantly denser than
the clumps in which they are embedded, and their free-fall times are
correspondingly shorter.  Insofar as the accretion time (and therefore
$t_w$) is tied to the free-fall time, as in inside-out collapse models
for star formation \citep{1977ApJ...214..488S},
it follows that the star formation occurs on a time scale shorter than the
free-fall time of the clump.  In this case, the effect of gravity on
the shell's motion, which is characterized by the factor $c_g$
(eq. \ref{eq:shelldeltaEscape}), is relatively minor:
\begin{equation}\label{eq:cg}
c_g = \left(\frac{9-3\krho}{8-3\krho}\right)^{1/2},
\end{equation}
which is of order unity for $\krho\leq 2$.  We shall generally adopt
$\krho = 1$ for clumps, which is similar to or somewhat shallower than
the clump structures inferred from observations \citep[for instance,
][ estimate $1<\krho\lesssim 2$ in Bok globules]{1991ApJ...381..474Y}.

	The final distinction between clumps and cores is that the
clumps are generally not isothermal, and the rms velocity is often
significantly greater than the thermal sound speed.  Since both cores
and star-forming clumps are gravitationally bound, it follows that the
escape velocity is generally greater for clumps than for cores.
Normalizing $\vesc$ to 2 $\kms$\ for clumps, the efficiency parameter
becomes
\begin{equation}
X=0.29\; \frac{\lnth}{\ln 200}\left(\frac{\vesc}{2\ \kms}\right)
	\left(\frac{40\ \kms}{f_wv_w}\right).
\label{eq:Xnumb}
\end{equation}
For $X=0.3$ and $f_w = 0.2$, we have $\mstarmax=0.27\Mcl$ from equation
(\ref{eq:mstarmax}).  Thus, for $\Mcl\simgt 500\, \msun$, a clump has
enough mass to make a star with a mass up to $\mstarmax\simgt
140\,\msun$, significantly above the upper mass cutoff in most star
clusters in the Galaxy, even those that are far more massive than the
clumps we are considering \citep{1995ApJ...454..151M}.  Thus, we do
not expect the formation of a single massive star to consume all the
gas in a star-forming clump.

We shall neglect the loss of mass due to the winds themselves
[$m_w(1-\phi_w)\ll \Mej$] for star formation within a clump.  This
approximation is justified below by the fact that $\Mej \gg m_w$ for
typical conditions ($\mstar \gg \mstaro$). Furthermore, the accuracy
of this approximation is improved somewhat by the fact that $\phi_w$
is typically about 0.5
in clumps \citep{myphd}. This implies that the error in $\varepsilon$
introduced by ignoring the contribution of $\mw$ to $\Mej$ is only a
couple percent when $\varepsilon \lesssim 50\%$, as we will find is
typically the case in regions without massive stars.

To estimate $\varepsilon$ for a star-forming clump, we assume that all
the star formation occurs at the center of the clump.  As we shall see
below, this apparently drastic approximation is in fact
reasonably accurate.  The critical angle for the escape of the
swept-up shell is
\begin{equation}
\muesc=\left[1-\left(\frac{\mstar-\mstaro}{X\Mcl}\right)\right]^{1/2}
\end{equation}
from equations (\ref{eq:esc}) and (\ref{eq:mstaro}),
provided, of course, that $\mstar\geq\mstaro$.
For $\mstarmax\gg \mstar\gg\mstaro$, as is typically
the case, we have $\muesc\simeq 1$ and
equation (\ref{eq:epsilonc}) for the star formation
efficiency simplifies to:
\begin{equation}
\frac{1}{\varepsilon}=1+1.72\;\frac{\ln 200}{\lnth}
	\left(\frac{2\ \kms}
	{\vesc}\right)\left(\frac{f_wv_w}{40\ \kms}\right).
\label{eq:epsilond}
\end{equation}
Because we do not expect the intrinsic properties of a protostellar
wind, $\theta_0$ and $\fw\vw$, to correlate with the escape velocity
of the protostellar clump, equation (\ref{eq:epsilond}) indicates that
the efficiency with which clusters of low-mass stars are formed is a
simple, increasing function of $\vesc$.  An escape velocity of order
$2~\kms$ leads to $\varepsilon \simeq 40\%$.

Equation (\ref{eq:epsilond}) assumes that the mass ejected by each
outflow can be tallied independently of other outflows.  This can be
verified by considering two outflows that occur simultaneously, which
maximizes their interaction. If the angle between their axes is
significantly greater than $\thesc$ for either, then the momentum
output from each is negligible within an angle $\thesc$ of the other's
axis. Then, $\Mej$ is given by equation (\ref{eq:Mej}) for each
outflow individually. Alternatively, consider the case in which the
outflows are completely aligned. They would then act as a single
outflow whose momentum (and wind mass, assuming identical values of
$v_w$) is the sum of its two components.  So long as the combined mass
of stars that forms simultaneously is still much less than $\mstarmax
\simeq \Mcl/3$, so that our assumption $1-\muesc \ll 1$ still holds,
equation (\ref{eq:Mej}) shows that $\Mej$ is then the sum of the
values it would take for each outflow individually. In either case,
and thus quite generally, the relative timing and orientation of
outflows is of little importance when estimating the mass they eject.

\subsection{Numerical Models: Off-Center, Ionization-Regulated Star 
Formation in Clusters} 
\label{S:numerics}

We wish to assess the accuracy of equation (\ref{eq:epsilond}), in
which we assumed that each star forms at the exact center of its
clump. In this section, we present calculations of $\varepsilon$ for
off-center star formation.  As above, we assume that the clump is
spherically symmetric with $\rhocl \propto r^{-\krho}$; we set
$\krho=1$.  We draw all stars from the same initial mass function
independently of their position within the clump, and orient their
outflow axes at random. The IMF is taken to be a Miller-Scalo
log-normal distribution, with a lower cutoff at $0.057~\Msun$, so that
$\mbarstar = 0.5~\Msun$. We take stars to be formed in the clump at a
local rate $\tgstar^{-1}\equiv \Mdotstar/\Mcl$ equal to the ambipolar
diffusion rate $\tAD^{-1}$, as in the photoionization-regulation model
of \cite{M89}. Because $\tAD\propto \tff$ if cosmic-ray ionization
prevails, we take the star formation rate to vary with the free-fall
rate in the region of high extinction. The ambipolar diffusion rate
decreases sharply due to far-ultraviolet ionization as the visual
extinction to the surface $\Avs$ falls below about four magnitudes. Although
this transition can be described analytically \citep[][]{MBM2000}, for
simplicity we shall take the transition to be sudden:
$\tgstar(r)^{-1}\propto \tff(r)^{-1}\propto\rho(r)^{1/2}$ if $\Avs>4$
and $\tgstar\e =0$ if $\Avs<4$.

For each star, a grid of solid angles is constructed around its
outflow axis. The wind momentum is distributed on this grid, whose
resolution is concentrated toward the axis for improved numerical
accuracy. The clump mass per solid angle is also computed, using the
distance from the newborn star to the center of the clump, and using
the angle between its axis and the radial direction. The ratio between
$d\pw/d\Omega$ and $d\Mcl/d\Omega$ gives the shell velocity at the
clump surface; this is compared to $c_g\vesc$, and the total ejected
mass is tallied. Thus, the effect of gravity is included as an overall
factor, rather than considered in the motion of each outflow element.

\begin{figure*}
\centerline{\epsfig{figure=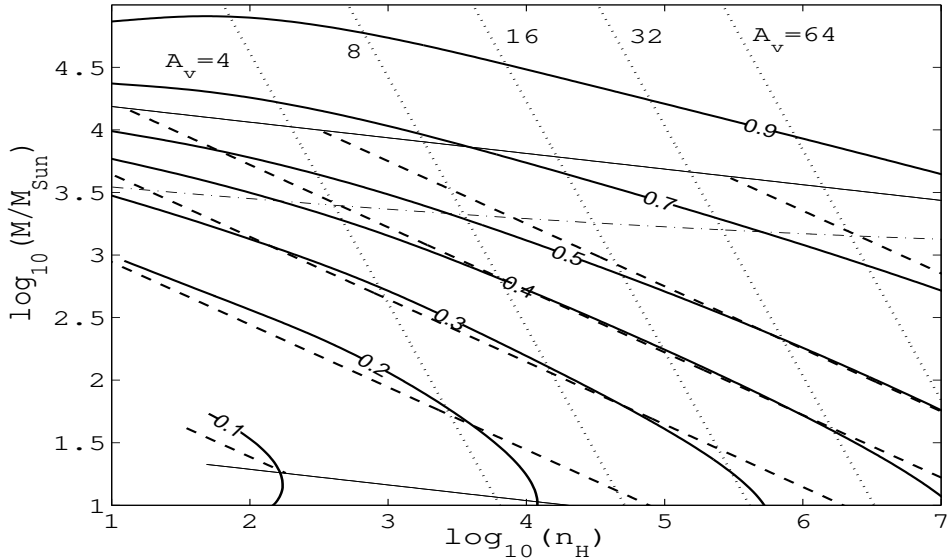, width=5in, height = 3in, angle=0}}
\caption[Analytical estimates of star-formation efficiency compared to
numerical simulation] {\small Estimates of the instantaneous star
formation efficiency $\varepsilon$ due to protostellar outflows. {\em
Dashed lines:} analytical estimate, equation (\ref{eq:epsilond}).
{\em Solid lines}: Numerical evaluation of the same quantity in a
protostellar clump with the structure $\rho\propto r^{-1}$ and gas
conversion rate scaling as $\rho^{1/2}$ in regions extinguished by
more than four magnitudes from the surface. In this calculation, the
effect of gravity on outflow motions has been included with an overall
correction factor $c_g$ (equation [\ref{eq:cg}]).  {\em Dotted
lines} indicate the mean extinction of the clump. Above the {\em
dash-dot line}, stars more massive than $20~\Msun$ are likely to form;
our estimates are therefore only rough upper limits in this region.
Below the {\em lower thin line}, $\mstarmax$ is less than twice the
mean stellar mass; above the {\em upper thin line}, $\mstaro$
exceeds the mean stellar mass. The analytical approximation requires
$\mstaro\ll\mbarstar\ll\mstarmax$, and breaks down near the thin
lines. \label{fig:clumpeffic} }
\end{figure*}

In Figure \ref{fig:clumpeffic}, we compare the results of this numerical
investigation to our analytical formulae. Allowing for distributed
star formation and averaging over the initial mass function changes
the estimated efficiencies appreciably only when the assumption
$\mstaro\ll\mbarstar\ll\mstarmax$ breaks down (near the thin lines
that represent these limits in Figure \ref{fig:clumpeffic}). For instance, the
curvature of the numerical contours of $\varepsilon$ away from the
analytical contours at the bottom of the plot can be attributed to the
smallness of $\mstarmax/\mbarstar$ in that region. We have also
compared to a calculation of $\varepsilon$ that averages over the
stellar initial mass function but assumes each star forms at the
center: this reproduces the off-center calculation almost exactly,
except for $\mstaro>\mbarstar$.  In general, assuming that each star
forms at the center yields an excellent approximation of the star
formation efficiency.

\subsection{Comparison with Observations}\label{S:ClumpObs}
The above theory implies that the clumps that give rise to clusters of
low-mass stars, like those in L1630 \citep[$n_H\sim 10^4~\cm^{-3}$,
$200<\Mcl<500~\Msun$;][]{L92}, form each star at an efficiency $30\%
\lesssim \varepsilon \lesssim 40\%$. This range of efficiencies agrees
with the theoretical arguments by \cite{1983ApJ...267L..97M} and
\cite{1983MNRAS.203.1011E}, who compared the properties of known
embedded and revealed bound clusters of similar masses and determined
that star formation efficiencies of about $30\%$ or $55\%$ would be 
necessary for the embedded bound clusters to evolve into 
the revealed bound clusters,
in the respective limits of slow or rapid
gas dispersal.

As described in \S \ref{S:efficdefs}, the observable efficiency is not
the instantaneous efficiency, $\varepsilon$, but instead the ratio of
stellar to total mass, $\SFE$. Whereas $\varepsilon$ is a simple
function of the current escape velocity from the system
(eq. \ref{eq:epsilond}), $\SFE$ is expected to vary from zero to unity
as stars form and gas is dissipated; $\varepsilon$ is only a rough
estimate for the characteristic value of $\SFE$ in the midst of this
process. Without knowing the evolution of the system, it is therefore
difficult to compare an observation of $\SFE$ to our prediction of
$\varepsilon$. In L1630, the observations of \cite{L92} imply
$\SFE\simeq 7\%$ (NGC~2023 and LBS~23), $19\%$ (NGC~2071), and $42\%$
(NGC~2024 and 2068) if one assumes each detected source signifies
$1~\Msun$ of stellar material. This assumption is relatively uncertain
due to variable extinction in the area, luminosity variation in the
embedded sources, and the completeness limit of the survey. If instead
each source signified the mean stellar mass ($\sim 0.5 ~\Msun$), the
inferred efficiencies would be about a factor of two lower. A direct
comparison with theory is therefore difficult for both theoretical and
experimental reasons, but it is encouraging that there is rough
agreement for the richer clusters, which may be more evolved.

\cite{1998ApJ...502..296O} have conducted a survey in Taurus for dense
gas and young stellar objects in various phases of evolution, in order
to ascertain the evolutionary sequence leading to cluster
formation. They find that only those clumps (and {\em all} those
clumps) with mean visual extinction $A_V>8~{\rm mag}$ are actively
forming stars, in agreement with the theoretical prediction of
\cite{M89}.  However, some clumps of lower extinction contain more
evolved objects that presumably formed while the column density was
above this threshold. Considering the decrease in column density to
correspond to a loss of gas mass, and multiplying by the mean area per
star, these authors estimate that about $4~\Msun$ is lost per observed
object, i.e., $\varepsilon({\rm obs}) \sim 11\%$ if each object
represents $0.5~\Msun$ of stellar material. This observational
determination of efficiency is also quite uncertain.  It should be
considered a lower limit for several reasons: the stellar sample is
likely to miss a significant fraction of weak-line T Tauri stars
\citep{1999AJ....118.1354B,1998AJ....115.2074B}; stars may have
drifted away from the older clumps; and the decrease in column density
could be partly due to an expansion of the clumps over time. Moreover,
the currently star-forming clumps have not created their full
complement of stars: on average, one might expect about half the stars
to have been born by now, in which case $\varepsilon({\rm obs}) \sim
22\%$.  The star-forming clumps in this sample have escape velocities
ranging from $0.8$ to $1.4~\kms$, for which we estimate
$\varepsilon\sim 25\pm 5\%$ from equation (\ref{eq:epsilond}). Again,
the theory predicts efficiencies similar to the values inferred from
observation.

\section{Conclusions}\label{S:conclusions}
The star formation efficiency is a fundamental property of the process
of star formation.  Star formation is often said to be inefficient
because it occurs much more slowly than it potentially could, as
discussed in the Introduction.  However, when defined in terms of the
fraction of the available mass that goes into stars, it is rather
high---at least for low-mass stars.  For such stars, the efficiency is
determined by the powerful protostellar winds that accompany the
formation of the star.  These winds produce a total momentum per unit
mass of star formed of order $m_wv_w/\mstar\equiv f_wv_w \simeq
40\,\kms$.  If this momentum were transferred to the surrounding gas
isotropically, then 
the mass ejected would be 
much greater than the mass that goes into the star, and the
efficiency would be 
low (Nakano et al 1995).  In fact, however, protostellar winds are
highly collimated.  To determine the star formation efficiency
resulting from collimated outflows, we assumed: (1) the interaction
between the protostellar wind and the ambient gas is mediated by thin,
radiative shocks that lead to radial outflow; (2) except possibly near
the wind axis, the mass of the wind is small compared with the mass
swept up by the wind shock; and (3) there is an angle $\thesc$
relative to the wind axis inside of which all mass is ejected and
outside of which mass can accrete
toward the central source, with a
fraction $f_w/(1+f_w)$ going into the protostellar wind.  Assumptions
(1) and (2) are consistent with observations of the momentum
distribution in protostellar outflows \citep{1999ApJ...526L.109M}.

We additionally assumed (4) that $f_w v_w$ is the same among stars of
different masses. This assumption, which is justified by the fact
that deuterium burning regulates protostars' escape velocity
\citep{1988ApJ...332..804S} and thus the wind velocity
\citep{1994ApJ...429..781S}, leads to an estimate of the star
formation efficiency that is the same for most stars and depends only
on the escape velocity of the region (eq. [\ref{eq:epsilond}]). If
this assumption did not hold, $\varepsilon$ would vary among stars.
However the conclusions of \S~\ref{S:Clumps} would still be valid, as
the star formation efficiency for a clump is an average over the IMF.

The principal effect of wind collimation on the momentum transfer to
the ambient gas is to concentrate the momentum in a mass that is
smaller than in the isotropic case by a factor of order
$1/\ln(2/\theta_0)$, where $\theta_0\sim 10\ee$ is the size of the
central region of the outflow (eq. \ref{eq:PmuHydromag}).  Gas is
ejected if its velocity exceeds the escape velocity, $\vesc$; as a
result, the mass ejected is given by $\Mej\vesc\sim \mstar f_w
v_w/\ln(2/\theta_0)$.  If the ejected mass is larger than the
protostellar mass, then the star formation efficiency is approximately
\begin{equation}
\varepsilon\simeq\frac{\mstar}{\Mej}\simeq
\frac{\vesc\ln(2/\theta_0)}{f_w v_w}.
\end{equation}
This result, which applies both to the formation of individual stars
in a core or the formation of a cluster of stars in a star-forming
clump, shows that the efficiency of formation of low-mass stars is
\begin{itemize}
\item[-] independent of the mass of the star; and

\item[-] only weakly dependent on the properties
of the core or clump, through $\vesc\propto 
(\Mcl/\Rcl)^{1/2}\propto \Mcl^{1/3}\rhocl^{1/6}$.
\end{itemize}
\noindent
Consequently, the efficiency is only weakly dependent on the
geometry of the star-forming region and the location of the outflow
source within it, as the results of \S\S \ref{S:coredisruption} and
\ref{S:numerics} demonstrate. 
These conclusions are in qualitative agreement with those of
\cite{1995ApJ...450..183N}, but our inclusion of the collimation of
the outflows increases the estimated star formation efficiency by
about an order of magnitude, so that $\varepsilon\sim 0.25-0.75$.

We have evaluated the star formation efficiency separately for
protostellar cores and for star-forming clumps.  Protostellar cores
are expected to be flattened in a direction perpendicular to the
outflow; since the outflow and inflow are in different directions, the
star formation efficiency is increased over the case of an isotropic
ambient medium.  We evaluated this effect quantitatively using the
Li-Shu (1996) model of isothermal, magnetized protostellar cores (see
Fig. 1).  Because $\varepsilon$ is independent of the stellar mass and
only weakly dependent on the core mass, we predict (in agreement with
Nakano et al 1995) that the initial mass function is just slightly
flatter than the mass distribution of protostellar cores, with a
power-law index $(0.9-1)\acore$.

Our model differs from most previous proposals in that the wind acts
to limit the fraction of the mass that can accrete, but does not cause
accretion to shut off entirely. This is consistent with the fact that
accretion is necessary to drive a powerful wind. Instead, accretion
terminates because only a finite reservoir of gas is available. In
contrast to the proposal of \cite{1998Natur.392..685V}, for instance,
the angle dividing accreted and ejected material takes a
characteristic value during the main accretion phase. Variations of
$\theta_{\rm esc}$ may result from a decline in the accretion rate,
but do not cause it. 
 
Another distinction of the theory presented here is that the division
between accretion and ejection occurs in swept-up material traveling
outward at the escape velocity -- material outside the region of
significant infall and far outside the centrifugal radius. Therefore
the complicated interaction among outflow, infall and rotation,
which determines the shape of the outflow cavity at its base, does not
affect the efficiency. All of the material at the base of the flow is
destined to accrete, for rotational flattening and infall only act to
lessen the effect of the wind.

We have found that the star formation efficiency is controlled by the
efficiency parameter $X$ (equation~[\ref{eq:defX}]), and in the case
of magnetically flattened protostellar cores, also by the flattening
parameter $(1+H_0)$. The factors that introduce uncertainty into the
theory are primarily those that affect $X$. For instance, the factor
$c_g$ that corrects for the effect of gravity on the outflow motions
is well-understood for clumps of hydrogen number density below
$10^5~\cm^{-3}$, in which winds are impulsive. As we discuss in \S
\ref{S:Wind-Structure-Duration}, we expect that winds are
approximately impulsive for higher mean densities as well; however,
the accuracy of this approximation is difficult to assess. For the
formation of an individual star, the wind duration is longer than the
free-fall time of the core. This raises the question, addressed in \S
\ref{S:G+Steady} of the Appendix, of what happens to material when it
reaches the core's surface.  The results presented in \S
\ref{S:coredisruption} presume that the shocked shell decouples from
the wind outside the core; however, equation
(\ref{eq:shellsteadyEscape}) shows that $c_g$ would be three times
lower in the opposite limit of perfect coupling.

A second source of uncertainty is the parameter $f_w v_w$ that relates
a star's final mass to the momentum of its wind. As discussed in \S
\ref{S:Wind-Structure-Duration}, our value, $f_w v_w=40 ~\kms$, is
consistent with current observational and theoretical estimates;
however, both are somewhat uncertain.

Our estimate of the effective broadening angle, $\theta_0$, of
protostellar winds is also not precise. \cite{1999ApJ...526L.109M}
estimate $\theta_0\simeq 10^{-2}$ by comparing the mass-velocity
distributions of outflows to the predictions of a simple model for
their dynamics, and by requiring that the model fit observations for
realistic inclinations of the wind axis to the line of sight. Note
that $\theta_0$ enters logarithmically in the efficiency parameter
$X$, as $\log(2/\theta_0)$; therefore, we consider it a minor source
of uncertainty as compared to $f_w v_w$. However, because the
intensity of protostellar winds along their axes is proportional to
$\theta_0^{-2}$, this parameter plays a more important role in
determining the minimum stellar mass $\mstaro$ capable of ejecting any
mass. The efficiency increases dramatically (and equation
[\ref{eq:epsilond}] breaks down) for clump masses large enough that
$\mstaro$ becomes comparable to the mean stellar mass.  For
$\theta_0=10^{-2}$, this happens when $\Mcl \gtrsim 10^4~\Msun$
(the upper thin line in figure \ref{fig:clumpeffic}). 
Another condition for the validity of our theory is that massive stars
do not dominate the ejection of mass; as shown by the dash-dot line in
that figure, we expect that to happen when $\Mcl \gtrsim
10^{3.5}~\Msun$. Thus for $\theta_0=10^{-2}$, the former is the more
restrictive criterion; if $\theta_0$ were greater than about
$10^{-1.5}$, the latter would be.

Lastly, we expect that the assumption of radial motion is not
perfect. Most likely, this implies that our estimate of $X$ should be
adjusted by a correction factor of order unity.  

For star-forming clumps, we found that the star-formation efficiency
is typically somewhat less than 0.5, although slight variations in the
parameters can bring $\varepsilon$ above 0.5.  As discussed in \S
\ref{S:Intro:clumps}, such high efficiencies might be expected to
leave the cluster gravitationally bound.  To show why most clusters
are unbound, we must consider the dynamics of cluster formation, a
topic we address in a future paper.

\acknowledgements 
We are grateful to Zhi-Yun Li, Scott Kenyon, and Lee Hartmann for
thought-provoking suggestions and comments, and to Ethan Vishniac for
discussing with us the stability of radiative shocks.  The research of
both CDM and CFM was supported in part by the National Science
Foundation through NSF grants AST 95-30480 and PHY94-07194, in part by
a NASA grant to the Center for Star Formation Studies; for CFM, in
part by a Guggenheim Fellowship; and for CDM, in part by an NSERC
fellowship.
CDM wishes to thank Roger Blandford and Sterl Phinney for their
generous hospitality during his visits to the California Institute of
Technology. 
CFM gratefully acknowledges the hospitality of John Bahcall of the
Institute for Advanced Study; his visit there was supported in part by
a grant from the Alfred P. Sloan Foundation.

\appendix 

\section{Effect of gravity on outflow motions} \label{S:GravAppdx}

Because $\vesc$ marks the boundary between ejected and retained
material, and because gravity is important for the dynamics of a shell
if its mean velocity is about $\vesc$, we wish to refine our estimate
of the escape criterion accounting for gravity.

Assume that a star forms at the center of the clump, and that the
clump's gravitational potential is not perturbed by the passage of the
outflow. Take the distribution of momentum injection to vary with
angle away from the outflow axis as
\begin{equation}
\frac{d\dot{p}_w}{d\Omega} = \frac{\dot{p}_w
	P(\mu)}{4\pi};
\end{equation} 
our model for $P(\mu)$ is given in equation (\ref{eq:dpwdOmega}).  We
take the distribution of clump density to be
\begin{equation}
\rhocl =\rhoo Q(\mu)r^{-\krho}
\end{equation} 
for $0<r<\Rcl$, where the normalization of the angular
distribution of the density is $\int_0^1 Q(\mu)d\mu=1$.
The mass per unit solid angle
enclosed within radius $r$ is therefore
\begin{equation}
\frac{d \Mcl}{d\Omega}= \frac{1}{4\pi} \Mcl(r)Q(\mu) = \frac{c_M}{4\pi}
	Q(\mu) r^{3-\krho},
\end{equation}
where the total mass inside
$r$ is $\Mcl(r)=c_M r^{3-\krho}$, and $c_M \equiv 4\pi \rhoo/(3-\krho)$.

	To determine the condition for the wind to eject material from
the surrounding clump or core, we make several approximations 
(\S \ref{S:mass}): We assume radial, momentum-conserving flow, and 
we neglect $dm_w/d\Omega$ compared with $d\Mcl/d\Omega$ in regions
in which the velocity of the shell is small compared with the wind
velocity, $v_s\ll v_w$.  In
addition, we adopt the monopole approximation for gravity, which is
generally quite good, even for flattened distributions
\citep[][]{1997ApJ...475..237L}.  The equation of motion for a sector
of the shell is then
\begin{equation}\label{eq:shellmotion}
\frac{d}{d t} \frac{d \Mcl}{d\Omega}v_s = \frac{d\dot{p}_w}{d\Omega} -
	\frac{GM}{r^2} \frac{dM}{d\Omega},  
\end{equation}
or 
\begin{equation}\label{eq:vmdot}
\frac{d}{d t} \Mcl(r)v_s =\frac{P(\mu)}{Q(\mu)}  {\dot{p}_w} -
	\frac{GM^2}{r^2},  
\end{equation}
where we have divided through by $Q(\mu)$. In the monopole
approximation, the kinematics of the shocked shell depend only on the
ratio $P(\mu)/Q(\mu)$, not on either one independently. 

If we neglect the effect of gravity on the propagation of the shock,
as in the main text, equation (\ref{eq:vmdot}) integrates to give
\begin{equation}\label{eq:mvs}
\Mcl v_s=\frac{P(\mu)}{Q(\mu)} p_w(t).
\end{equation}
Usually, we shall adopt the criterion that a sector of the shell
will escape from the clump if its velocity exceeds the escape
velocity.  The critical angle at which escape is marginally
possible, $\thesc=\cos\e \muesc$, 
is then determined by setting $p_w(t)=p_w$,
the total wind momentum, and $v_s=\vesc$ in equation
(\ref{eq:mvs}).  Including the effect of gravity 
on the dynamics of the shell increases
the required momentum by a factor $c_g\geq 1$ that we shall
determine below:
\begin{equation}
\frac{P(\muesc)}{Q(\muesc)}p_w =c_g \Mcl\vesc.
\label{eq:PQp}
\end{equation}

   To solve for the dynamics of the shock in the clump including
the effects of gravity, we
express $\Mcl(r)$ as $c_M r^{3-\krho}$ and $d/dt$
as $v_s d/dr$.  Equation (\ref{eq:shellmotion}) then becomes
\begin{equation}\label{eq:shellmotion2} 
c_M v_s \frac{d}{dr} v_s r^{3-\krho} = \frac{P(\mu)}{Q(\mu)}\dot{p}_w
    - G c_M^2 r^{4-\krho}. 
\end{equation}
We now introduce the variables $x\equiv r^{4-\krho}$ and $y\equiv
r^{3-\krho} v_s$; in terms of these variables, equation
(\ref{eq:shellmotion2}) becomes
\begin{equation} 
\frac{4-\krho}{2} c_M \frac{d y^2}{d x} =
	\frac{P(\mu)}{Q(\mu)}\dot{p}_w - G c_M^2 x^{(4-2\krho)/(4-\krho)}.
\end{equation}
Integrating, and transforming back to radius and velocity from $x$ and
$y$, we find
\begin{equation} \label{eq:shellmotion3}
\Mcl v_s^2 = \frac{2}{4-\krho}  \frac{P(\mu)}{Q(\mu)}\dot{p}_w r - 
	\frac{\Mcl\vesc(r)^2}{8-3\krho} + \frac{C}{\Mcl},
\end{equation} 
where we have defined $\vesc(r)\equiv \sqrt{2G\Mcl(r)/r}$ as the local
escape velocity [so that $\vesc(\Rcl)=\vesc$], and $C$ is a constant of
integration.

\subsection{Impulsive Winds} \label{S:G+Impulse}
Consider the case that the wind momentum is injected
impulsively, i.e, $\tw\ll\tff$. Then, $\dot{p}_w = 0$ during the
shell's expansion, so that (multiplying through by $\Mcl$)
\begin{equation} \label{eq:shelldeltamotion}
\Mcl^2 v_s^2 = - \frac{\Mcl^2 \vesc(r)^2}{8-3\krho} + C. 
\end{equation}
Evaluating this equation at $r=0$, where $\Mcl\vesc(r)=0$, we 
find $\Mcl Q(\mu) v_s= p_w P(\mu)$; therefore 
$C = [p_w P(\mu)/Q(\mu)]^2$, and
\begin{equation}\label{eq:shelldeltamotion2}
\Mcl^2 v_s^2 =  \left[\frac{P(\mu)}{Q(\mu)} p_w\right]^2 
	- \frac{\Mcl^2 \vesc(r)^2}{8-3\krho}.
\end{equation}
The condition for the escape is $v_s(\Rcl)>\vesc$, or
\begin{equation}\label{eq:shelldeltaEscape}
\frac{P(\mu)}{Q(\mu)} p_w > \left(\frac{9-3\krho}
	{8-3\krho}\right)^{1/2} \Mcl \vesc.
\end{equation}
Thus, in this case the factor by which the critical wind
force must be enhanced to offset the effects of gravity 
on the dynamics of the shell is 
$c_g=[(9-3\krho)/(8-3\krho)]^{1/2}$ (cf eq. \ref{eq:PQp}).  

\subsection{Steady Winds} \label{S:G+Steady}
Next, consider what happens if the wind momentum injection is constant
throughout the expansion of the shell, so that $\dot{p}_w$ maintains
the steady value $\pw/\tw$. Furthermore, let us suppose that the wind
continues to inject momentum after the shell breaks out of the
clump. Then, a na\"{\i}ve estimate for the criterion for escape is
that the wind force overcome gravity at the clump surface,
\begin{equation}\label{eq:shellsteadyestimate} 
\frac{P(\mu)}{Q(\mu)}p_w  \simgt \frac{G \Mcl^2 \tw}{\Rcl^2} = 
	\frac{\pi\tw}{4\tff} \Mcl\vesc. 
\end{equation}
This estimate gives $c_g=(\pi/4)(t_w/\tff)$ for the
factor by which the momentum in the wind must be enhanced
to overcome the effects of gravity on the shell dynamics.
This factor is large when $t_w\gg\tff$ since only
the momentum delivered in a time of order $\tff$ is
effective in ejecting the shell. In fact, the condition
$c_g\geq 1$ sets a lower bound
on the value of 
$t_w/\tff$ 
for a steady wind.

Consider the dynamics of the outflow shell after it has broken free
of the clump, so that its mass is constant. Then, since we are
assuming that $\pdotw$ is constant in time, we can integrate equation
(\ref{eq:vmdot}) directly, using $d/dt = v_s d/dr$:
\begin{equation}\label{eq:ShellOutsideClump} 
E(r) - E(\Rcl) = \frac{P(\mu)}{Q(\mu)}\pdotw (r-\Rcl) , 
\end{equation} 
where $E(r)=\Mcl v_s^2 /2 - G \Mcl^2/r$ is the energy of the shell and
$E(\Rcl)$ is its energy at the clump surface. If the wind were no
longer blowing, the shell's energy would be constant and positive
energy ($v_s^2>G\Mcl^2/r$) would lead to escape. However, we are
assuming that the wind continues to blow well past the time that the
shocked shell reaches the clump surface.  In this case, the shell
might escape even if it has negative energy, because of the wind
force.  (Below, we shall consider the case that the outflow decouples
from the wind after breaking out.)

We shall evaluate $v_s(\Rcl)$ using equation (\ref{eq:shellmotion3}), and
then use equation (\ref{eq:ShellOutsideClump}) to determine the
criterion for escape.  We readily see that $C=0$ for a steady wind, for all
the other terms in equation (\ref{eq:shellmotion3}) are zero at the
origin; therefore, 
\begin{eqnarray} \label{eq:shellsteadymotion} 
\Mcl v_s^2 = \frac{2}{4-\krho}\frac{\pw r P(\mu)}{\tw Q(\mu)} - \frac{\Mcl
	\vesc(r)^2}{8-3\krho}  ~~~~~ { (r\leq \Rcl) }.
\end{eqnarray}
Using the value of $E(\Rcl)$ from this in equation
(\ref{eq:ShellOutsideClump}), and requiring that $dE(r) > -G
\Mcl d\Mcl/r$ [so
that $v_s^2(r)>0$] for all $r$, we find the critical value of the wind
force: 
\begin{equation}\label{eq:shellsteadyEscape}
\frac{P(\mu)}{Q(\mu)}p_w  >\left(\frac{4-\krho}{3-\krho}\right)^2 
\left\{1- \left[\frac{5-2\krho}{(8-3\krho)(4-\krho)}
\right]^{1/2}  \right\}^2 
\times
\frac{\pi\tw}{4\tff} \Mcl\vesc. 
\end{equation}
This more precise wind force criterion is $35\%$ and $31\%$ lower than
the simple estimate of equation (\ref{eq:shellsteadyestimate}) for
$\krho = 0$ and $\krho = 1$, respectively. The two agree exactly for
$\krho=2$, because gravity and wind force scale in the same manner
inside the clump in that case. This implies that the shell velocity at
the surface vanishes if the wind force equals the critical value if
$\krho=2$, exactly as we assumed in estimate
(\ref{eq:shellsteadyestimate}).

Equations (\ref{eq:shellsteadyestimate}) --
(\ref{eq:shellsteadyEscape}) assume that the wind continues to
deposit momentum into the shell efficiently, even after the outflow
has emerged from the clump. There are two reasons that this assumption
may be incorrect, which make the above estimates lower limits for the
critical wind force. First, if $\tw$ is not too much longer than
$\tff$, there is the possibility that the wind force will taper off
before the shell has achieved escape velocity. Second, there is the
possibility that the shell will be disrupted by a thin-shell
instability. The former effect can be ignored, because it is only
important when $\tw\simeq\tff$. The latter is more serious, as it
would remove the supporting effect of the wind (partially or
completely) once the instability set in. 

Shock-bounded thin shells are subject to nonlinear instabilities
\citep[e.g.,][]{1994ApJ...428..186V}; these are expected to become
stronger if the dense fluid (the ambient gas, in this case) is
``upward'', i.e., if the shell accelerates, or at least decelerates
more slowly than gravity would predict (E. T. Vishniac, private
communication).  Writing out $d(d\Mcl/d\Omega)/dt$ as $r^2 v_s \rhocl$ and
$d\dot{p}_w/d\Omega$ as $\rhow \vw^2 r^2$, we find from equation
(\ref{eq:shellmotion}) that the difference between the actual
acceleration and gravity is
\begin{equation}\label{eq:RTdiscriminant}
\frac{dv_s}{dt} + \frac{G \Mcl}{r^2} = \frac{r^2}{d\Mcl/d\Omega} (\rhow
\vw^2 - \rhocl v_s^2). 
\end{equation}
Since gravity is an additive term in the acceleration, we see that the
shell decelerates more slowly than gravity would predict (and is
unstable) if the same shell would accelerate, were gravity suddenly
turned off. Equation (\ref{eq:RTdiscriminant}) shows that this only
occurs if the ram pressure in the wind exceeds the ram pressure of the
ambient gas.  For a steady wind, this requires that the ambient medium
must steepen so that $\krho>2$ (declining winds would require even
steeper profiles).  We therefore expect that shells might become
partially decoupled from their winds as they emerge from a clump, not
before.

Although this instability is nonlinear, and may be suppressed by
magnetic fields in the outflow shell or in the wind, we can obtain an
upper limit to the critical wind force if we assume that outflows
decouple entirely from their winds after breakout. Then, the criterion
for ejection is once again $v_s(\Rcl) > \vesc$. From equation
(\ref{eq:shellsteadymotion}), this requires
\begin{equation}\label{eq:shellsteadyEscape2}
\frac{P(\mu)}{Q(\mu)}p_w  >  (4-\krho) \frac{9-3\krho}{8-3\krho}
\times \frac{G \Mcl^2 \tw}{\Rcl^2} = (4-\krho) \frac{9-3\krho}{8-3\krho}
\times \frac{\pi\tw}{4\tff} \Mcl \vesc. 
\end{equation}
The factor $c_g$ in this case is the coefficient of
$\Mcl\vesc$ in the equation above.
We see that the wind force required to eject the shell is much
stronger if the shell decouples from the wind after breakout than if
the wind continues to provide support, the factor between equations
(\ref{eq:shellsteadyEscape}) and (\ref{eq:shellsteadyEscape2}) being
seven or three if $\krho$ is zero or two, respectively.


\begin{thebibliography}{62}
\expandafter\ifx\csname natexlab\endcsname\relax\def\natexlab#1{#1}\fi

\bibitem[{{Adams} \& {Fatuzzo}(1996)}]{1996ApJ...464..256A}
{Adams}, F.~C. \& {Fatuzzo}, M. 1996, \apj, 464, 256

\bibitem[{{Adams} {et~al.}(1989){Adams}, {Ruden}, \&
  {Shu}}]{1989ApJ...347..959A}
{Adams}, F.~C., {Ruden}, S.~P., \& {Shu}, F.~H. 1989, \apj, 347, 959

\bibitem[{{Arce} \& {Goodman}(2000)}]{AG99}
{Arce}, H. \& {Goodman}, A. 2000, in prep.

\bibitem[{{Bally} \& {Duvert}(1994)}]{BCD94}
{Bally}, J.and~{Castets}, A. \& {Duvert}, G. 1994, \apj, 423, 310

\bibitem[{{Bally} {et~al.}(1999){Bally}, {Reipurth}, {Lada}, \&
  {Billawala}}]{BRLB99}
{Bally}, J., {Reipurth}, B., {Lada}, C.~J., \& {Billawala}, Y. 1999, \aj, 117,
  410

\bibitem[{{Benson} \& {Myers}(1989)}]{1989ApJS...71...89B}
{Benson}, P.~J. \& {Myers}, P.~C. 1989, \apjs, 71, 89

\bibitem[{{Bertoldi} \& {McKee}(1996)}]{BM96}
{Bertoldi}, F. \& {McKee}, C.~F. Self-regulated star formation in molecular
  clouds (Amazing Light: A Volume Dedicated to C.H. Townes on his 80th
  Birthday, ed. R.Y.Chiao (New York: Springer)), 41

\bibitem[{{Bontemps} {et~al.}(1996){Bontemps}, {Andre}, {Terebey}, \&
  {Cabrit}}]{1996A&A...311..858B}
{Bontemps}, S., {Andre}, P., {Terebey}, S., \& {Cabrit}, S. 1996, \aap, 311,
  858

\bibitem[{{Brice{\~n}o} {et~al.}(1999){Brice{\~n}o}, {Calvet}, {Kenyon}, \&
  {Hartmann}}]{1999AJ....118.1354B}
{Brice{\~n}o}, C., {Calvet}, N., {Kenyon}, S., \& {Hartmann}, L. 1999, \aj,
  118, 1354

\bibitem[{{Brice{\~n}o} {et~al.}(1998){Brice{\~n}o}, {Hartmann}, {Stauffer}, \&
  {Mart{\'i}n}}]{1998AJ....115.2074B}
{Brice{\~n}o}, C., {Hartmann}, L., {Stauffer}, J., \& {Mart{\'i}n}, E. 1998,
  \aj, 115, 2074

\bibitem[{{Caselli} \& {Myers}(1995)}]{1995ApJ...446..665C}
{Caselli}, P. \& {Myers}, P.~C. 1995, \apj, 446, 665

\bibitem[{{Elmegreen}(1983)}]{1983MNRAS.203.1011E}
{Elmegreen}, B.~G. 1983, \mnras, 203, 1011

\bibitem[{{Goldsmith} {et~al.}(1986){Goldsmith}, {Langer}, \&
  {Wilson}}]{1986ApJ...303L..11G}
{Goldsmith}, P.~F., {Langer}, W.~D., \& {Wilson}, R.~W. 1986, \apjl, 303, L11

\bibitem[{{Hills}(1980)}]{1980ApJ...235..986H}
{Hills}, J.~G. 1980, \apj, 235, 986

\bibitem[{{Lada} \& {Gautier}(1982)}]{LG82}
{Lada}, C.~J. \& {Gautier}, T.~N., I. 1982, \apj, 261, 161

\bibitem[{{Lada} {et~al.}(1984){Lada}, {Margulis}, \&
  {Dearborn}}]{1984ApJ...285..141L}
{Lada}, C.~J., {Margulis}, M., \& {Dearborn}, D. 1984, \apj, 285, 141

\bibitem[{{Lada}(1992)}]{L92}
{Lada}, E.~A. 1992, \apjl, 393, L25

\bibitem[{{Lada} {et~al.}(1991){Lada}, {Evans}, {Depoy}, \& {Gatley}}]{LEDG91}
{Lada}, E.~A., {Evans}, Neal~J., I., {Depoy}, D.~L., \& {Gatley}, I. 1991,
  \apj, 371, 171

\bibitem[{{Ladd} {et~al.}(1998){Ladd}, {Fuller}, \&
  {Deane}}]{1998ApJ...495..871L}
{Ladd}, E.~F., {Fuller}, G.~A., \& {Deane}, J.~R. 1998, \apj, 495, 871

\bibitem[{{Langer} {et~al.}(1986){Langer}, {Frerking}, \&
  {Wilson}}]{1986ApJ...306L..29L}
{Langer}, W.~D., {Frerking}, M.~A., \& {Wilson}, R.~W. 1986, \apjl, 306, L29

\bibitem[{{Larson}(1981)}]{1981MNRAS.194..809L}
{Larson}, R.~B. 1981, \mnras, 194, 809

\bibitem[{{Lee} {et~al.}(1999){Lee}, {Myers}, \&
  {Tafalla}}]{1999ApJ...526..788L}
{Lee}, C.~W., {Myers}, P.~C., \& {Tafalla}, M. 1999, \apj, 526, 788

\bibitem[{{Levreault}(1984)}]{1984ApJ...277..634L}
{Levreault}, R.~M. 1984, \apj, 277, 634

\bibitem[{{Li} {et~al.}(1997){Li}, {Evans}, \& {Lada}}]{1997ApJ...488..277L}
{Li}, W., {Evans}, Neal~J., I., \& {Lada}, E.~A. 1997, \apj, 488, 277

\bibitem[{{Li}(1998)}]{1998ApJ...497..850L}
{Li}, Z.-Y. 1998, \apj, 497, 850

\bibitem[{{Li} \& {Shu}(1996)}]{1996ApJ...472..211L}
{Li}, Z.-Y. \& {Shu}, F.~H. 1996, \apj, 472, 211

\bibitem[{{Li} \& {Shu}(1997)}]{1997ApJ...475..237L}
{Li}, Z.-Y. \& {Shu}, F.~H. 1997, \apj, 475, 237

\bibitem[{{Massey} {et~al.}(1995){Massey}, {Johnson}, \&
  {DeGioia-Eastwood}}]{1995ApJ...454..151M}
{Massey}, P., {Johnson}, K., \& {DeGioia-Eastwood}, K. 1995, \apj, 454, 151

\bibitem[{{Mathieu}(1983)}]{1983ApJ...267L..97M}
{Mathieu}, R.~D. 1983, \apjl, 267, L97

\bibitem[{{Matzner} {et~al.}(2000){Matzner}, {Bertoldi}, \& {McKee}}]{MBM2000}
{Matzner}, C., {Bertoldi}, F., \& {McKee}, C.~F. 2000, in prep.

\bibitem[{{Matzner}(1999)}]{myphd}
{Matzner}, C.~D. 1999, PhD thesis, U.~C. Berkeley

\bibitem[{{Matzner} \& {McKee}(1999)}]{1999ApJ...526L.109M}
{Matzner}, C.~D. \& {McKee}, C.~F. 1999, \apjl, 526, L109

\bibitem[{{McKee}(1989)}]{M89}
{McKee}, C.~F. 1989, \apj, 345, 782

\bibitem[{{McKee} \& {Williams}(1997)}]{1997ApJ...476..144M}
{McKee}, C.~F. \& {Williams}, J.~P. 1997, \apj, 476, 144

\bibitem[{{Miller} \& {Scalo}(1978)}]{MS78}
{Miller}, G.~E. \& {Scalo}, J.~M. 1978, \pasp, 90, 506

\bibitem[{{Momose} {et~al.}(1996){Momose}, {Ohashi}, {Kawabe}, {Hayashi}, \&
  {Nakano}}]{1996ApJ...470.1001M}
{Momose}, M., {Ohashi}, N., {Kawabe}, R., {Hayashi}, M., \& {Nakano}, T. 1996,
  \apj, 470, 1001

\bibitem[{{Motte} {et~al.}(1998){Motte}, {Andr\'e}, \&
  {Neri}}]{1998A&A...336..150M}
{Motte}, F., {Andr\'e}, P., \& {Neri}, R. 1998, \aap, 336, 150

\bibitem[{{Mouschovias}(1991)}]{1991ApJ...373..169M}
{Mouschovias}, T.~C. 1991, \apj, 373, 169

\bibitem[{{Myers} {et~al.}(1988){Myers}, {Heyer}, {Snell}, \&
  {Goldsmith}}]{1988ApJ...324..907M}
{Myers}, P.~C., {Heyer}, M., {Snell}, R.~L., \& {Goldsmith}, P.~F. 1988, \apj,
  324, 907

\bibitem[{{Najita} \& {Shu}(1994)}]{1994ApJ...429..808N}
{Najita}, J.~R. \& {Shu}, F.~H. 1994, \apj, 429, 808

\bibitem[{{Nakano} {et~al.}(1995){Nakano}, {Hasegawa}, \&
  {Norman}}]{1995ApJ...450..183N}
{Nakano}, T., {Hasegawa}, T., \& {Norman}, C. 1995, \apj, 450, 183

\bibitem[{{Norman} \& {Silk}(1980)}]{NS80}
{Norman}, C. \& {Silk}, J. 1980, \apj, 238, 158

\bibitem[{{Onishi} {et~al.}(1998){Onishi}, {Mizuno}, {Kawamura}, {Ogawa}, \&
  {Fukui}}]{1998ApJ...502..296O}
{Onishi}, T., {Mizuno}, A., {Kawamura}, A., {Ogawa}, H., \& {Fukui}, Y. 1998,
  \apj, 502, 296

\bibitem[{{Pelletier} \& {Pudritz}(1992)}]{1992ApJ...394..117P}
{Pelletier}, G. \& {Pudritz}, R.~E. 1992, \apj, 394, 117

\bibitem[{{Reipurth} {et~al.}(1997){Reipurth}, {Bally}, \&
  {Devine}}]{1997AJ....114.2708R}
{Reipurth}, B., {Bally}, J., \& {Devine}, D. 1997, \aj, 114, 2708

\bibitem[{{Safier} {et~al.}(1997){Safier}, {McKee}, \&
  {Stahler}}]{1997ApJ...485..660S}
{Safier}, P.~N., {McKee}, C.~F., \& {Stahler}, S.~W. 1997, \apj, 485, 660

\bibitem[{{Shu} {et~al.}(1994){Shu}, {Najita}, {Ostriker}, {Wilkin}, {Ruden},
  \& {Lizano}}]{1994ApJ...429..781S}
{Shu}, F., {Najita}, J., {Ostriker}, E., {Wilkin}, F., {Ruden}, S., \&
  {Lizano}, S. 1994, \apj, 429, 781

\bibitem[{{Shu}(1977)}]{1977ApJ...214..488S}
{Shu}, F.~H. 1977, \apj, 214, 488

\bibitem[{{Shu} {et~al.}(1988){Shu}, {Lizano}, {Ruden}, \&
  {Najita}}]{1988ApJ...328L..19S}
{Shu}, F.~H., {Lizano}, S., {Ruden}, S.~P., \& {Najita}, J. 1988, \apjl, 328,
  L19

\bibitem[{{Shu} {et~al.}(1995){Shu}, {Najita}, {Ostriker}, \&
  {Shang}}]{1995ApJ...455L.155S}
{Shu}, F.~H., {Najita}, J., {Ostriker}, E.~C., \& {Shang}, H. 1995, \apjl, 455,
  L155

\bibitem[{{Shu} {et~al.}(1991){Shu}, {Ruden}, {Lada}, \& {Lizano}}]{Shuea91}
{Shu}, F.~H., {Ruden}, S.~P., {Lada}, C.~J., \& {Lizano}, S. 1991, \apjl, 370,
  L31

\bibitem[{{Shu} {et~al.}(1990){Shu}, {Tremaine}, {Adams}, \&
  {Ruden}}]{1990ApJ...358..495S}
{Shu}, F.~H., {Tremaine}, S., {Adams}, F.~C., \& {Ruden}, S.~P. 1990, \apj,
  358, 495

\bibitem[{{Silk}(1995)}]{1995ApJ...438L..41S}
{Silk}, J. 1995, \apjl, 438, L41

\bibitem[{{Stahler}(1988)}]{1988ApJ...332..804S}
{Stahler}, S.~W. 1988, \apj, 332, 804

\bibitem[{{Terebey} {et~al.}(1984){Terebey}, {Shu}, \&
  {Cassen}}]{1984ApJ...286..529T}
{Terebey}, S., {Shu}, F.~H., \& {Cassen}, P. 1984, \apj, 286, 529

\bibitem[{{Testi} \& {Sargent}(1998)}]{1998ApJ...508L..91T}
{Testi}, L. \& {Sargent}, A.~I. 1998, \apjl, 508, L91

\bibitem[{{Umemoto} {et~al.}(1991){Umemoto}, {Hirano}, {Kameya}, {Fukui},
  {Kuno}, \& {Takakubo}}]{1991ApJ...377..510U}
{Umemoto}, T., {Hirano}, N., {Kameya}, O., {Fukui}, Y., {Kuno}, N., \&
  {Takakubo}, K. 1991, \apj, 377, 510

\bibitem[{{Velusamy} \& {Langer}(1998)}]{1998Natur.392..685V}
{Velusamy}, T. \& {Langer}, W.~D. 1998, \nat, 392, 685

\bibitem[{{Vishniac}(1994)}]{1994ApJ...428..186V}
{Vishniac}, E.~T. 1994, \apj, 428, 186

\bibitem[{{Williams} {et~al.}(2000){Williams}, {Blitz}, \& {McKee}}]{WBM}
{Williams}, J.~P., {Blitz}, L., \& {McKee}, C.~F. The Structure and Evolution
  of Molecular Clouds: From Clumps to Cores to the IMF (University of Arizona
  Press), in press

\bibitem[{{Yun} \& {Clemens}(1991)}]{1991ApJ...381..474Y}
{Yun}, J.~L. \& {Clemens}, D.~P. 1991, \apj, 381, 474

\bibitem[{{Zuckerman} \& {Palmer}(1974)}]{1974ARA&A..12..279Z}
{Zuckerman}, B. \& {Palmer}, P. 1974, \araa, 12, 279

\end{thebibliography}

\end{document}